\documentclass[preprint,prc,aps,eqsecnum,amssymb,floatfix]{revtex4}
\usepackage{bm}
\usepackage{graphicx}

\newcommand{\bfr}{{\bm r}}

\newcommand{\bfk}{{\bm k}}

\newcommand{\bfp}{{\bm p}}

\newcommand{\bfx}{{\bm x}}
\newcommand{\bfy}{{\bm y}}

\begin{document}
%
\title{$N-d$ Elastic Scattering Using the Hyperspherical Harmonics Approach 
with Realistic Local and Non-Local Interactions}

\author{L.E. Marcucci$^{1,2}$, A. Kievsky$^2$, L. Girlanda$^2$,
S. Rosati$^{1,2}$, and M. Viviani$^2$}

\affiliation{
$^1$ Department of Physics ``E. Fermi'', University of Pisa, 
Largo B. Pontecorvo 3, I-56127, Pisa, Italy \\
$^2$  Istituto Nazionale di Fisica Nucleare, Sezione di Pisa, 
Largo B. Pontecorvo 3, I-56127, Pisa, Italy \\ }

\begin{abstract}

The application of the hyperspherical harmonic approach to the 
case of the $N-d$ scattering problem below deuteron breakup threshold 
is described. 
The nuclear Hamiltonian includes two- and three-nucleon 
interactions, in particular 
the Argonne $v_{18}$, the N3LO-Idaho, and the $V_{low\!-\!k}$ two-nucleon, 
and the Urbana IX and N2LO three-nucleon interactions. 
Some of these models are local, some are non-local. Also 
electromagnetic effects are included. 
Accurate calculations for 
many scattering observables at various center-of-mass energies 
are performed and the results are compared with the 
available experimental data. Furthermore, a 
$\chi^2$ analysis of some of the Hamiltonian models has been performed
to compare their capability to describe the scattering process.

\end{abstract}
\maketitle

\section{Introduction}
\label{sec:intro}

One of the main inputs for any study on nuclear 
systems within a non-relativistic framework is the model used  
to describe the nuclear interaction, i.e. the nuclear Hamiltonian. 
Nowadays, it is common practice to use, at least for 
few-nucleon systems, Hamiltonian models composed of a 
two-nucleon plus, for $A\geq 3$, a three-nucleon interaction (TNI).
The modern two-nucleon interaction models have a large number 
of parameters and 
can reproduce the deuteron properties and the nucleon-nucleon 
scattering data up to the pion threshold with a $\chi^2$/datum $\simeq 1$. 
Among them, the Argonne $v_{18}$ (AV18)~\cite{Wir95} and 
the charge-dependent Bonn (CDBonn)~\cite{Mac01} explicitly 
include charge-symmetry-breaking terms in the nuclear 
interaction, in order to reproduce equally well the $np$ and $pp$ data.
Recently, a number of two-nucleon interaction models have been 
derived by many authors within an effective field theory (EFT)
approach, up to next-to-next-to-next-to leading 
order (N3LO)~\cite{Epe98,Ent03}. In particular, the N3LO model 
of Ref.~\cite{Ent03} (N3LO-Idaho) reaches the same level of 
accuracy of the CDBonn model.

The available models for the TNI contain, on the contrary to the two-nucleon 
interaction models, a small number 
of parameters, usually fixed to reproduce the $^3$H and/or $^4$He 
binding energies
and, in some cases, the nuclear matter equilibrium density.
Among the different existing models, we quote only those ones 
of the Urbana and Tucson-Melbourne families. 
Although constructed within different frameworks, these 
two families of potentials have shown to give similar results, 
when used in conjunction with a given two-nucleon interaction model.
Therefore, we have considered the Urbana IX~\cite{Pud95} (UIX) 
TNI in conjunction with both the AV18 and N3LO-Idaho two-nucleon 
interaction  models. Finally, it should be noticed that within the 
EFT approach mentioned above, also TNIs appear at the 
next-to-next-to leading order (N2LO)~\cite{Epe02}. 
In particular, we will consider 
the local version of this N2LO TNI, as given in Ref.~\cite{Nav07}.

More recently, a new class of two-nucleon interactions has been 
obtained ($V_{low-k}$ potentials). With the purpose of 
eliminating from the semi-phenomenological high precision two-nucleon 
potentials the high-momentum parts,  the two-nucleon 
Hilbert space has been separated 
into low- and high-momentum regions and the renormalization group
method has been used to integrate out the high-momentum components 
above a cutoff $\Lambda$~\cite{Bog07}. 
The value for $\Lambda$ is typically fixed in $A>2$ systems, for 
example so 
that the triton binding energy is reproduced.

At this point, a crucial issue is to test the model for 
the nuclear Hamiltonian studying $A\geq 4$ bound states and $A\geq 3$
scattering states. In the present work, we focus our attention to the $A=3$
scattering problem, which has been the object of a large number of
investigations~\cite{review}. 
Traditionally, the $A=3$ scattering problem with
realistic Hamiltonians is solved using the Faddeev equations. 
On the other hand, 
we have developed in recent years a variational approach, based on the 
expansion of the wave functions on the hyperspherical harmonics (HH) basis
(for a recent review, see Ref.~\cite{Kie08}). This method has
proven to be very efficient in the description of bound and
scattering states in few-nucleon systems. In Ref.~\cite{Kie93}
the HH expansion with correlations factors (the 
correlated HH -- CHH -- and 
pair-correlated HH -- PHH -- expansions)
has been used to describe $A=3$ bound states, whereas the extension
to scattering states has been discussed in Ref.~\cite{Kie94}. The
inclusion of correlation factors was motivated by the short
range repulsion of the two-nucleon potential which induces particular
configurations in the wave function difficult to describe using the
bare expansion. In fact, in Ref.~\cite{Kie97a} the HH expansion
without correlation factors has been used to describe the $A=3$
bound state, with the AV18 interaction. The conclusion was that
a much higher number of states are necessary when the bare
expansion is used. The same observation has been done in the $A=4$
system~\cite{Viv05} and is a direct consequence of using local
interactions, which result to have a strong repulsion at short
distances. The implementation of the HH method in momentum-space
has been done in Ref.~\cite{Viv06} for the $A=3,4$ bound states.
This analysis has revealed a much faster convergence of the
expansion when non-local potentials are considered, even when
TNI terms are taken into account. 

The aim of the present work is twofold. First, we want to extend the HH
method to describe $N-d$ scattering states using either local or non-local
potentials. We will show that we can apply the method in both configuration
and momentum spaces. Second, we will present a detailed
comparison between the predictions of the different
models, local and non-local, at low center-of-mass energies,
for $n-d$ as well as $p-d$ scattering. Moreover, we will consider
the Coulomb potential plus the magnetic moment (MM) interaction that
was shown to give sizable contributions~\cite{Kie04}.
To our knowledge, this is the
first time that non-local two- plus three-nucleon potentials 
are used to describe $p-d$ scattering at very low energies.

The paper is organized as follows: in Sec.~\ref{sec:form}, the 
HH method for the low-energy scattering problem is 
described, putting more emphasis on those new developments of 
the method necessary in order to use non-local interaction models. In 
Sec.~\ref{sec:res}, the results for the zero-energy scattering lengths and 
low-energy elastic scattering observables are
presented and discussed. Some concluding remarks are given in 
Sec.~\ref{sec:concl}.

\section{Formalism}
\label{sec:form}

In this section we present the HH method for scattering states. 
The method for bound states has been 
most recently reviewed in Ref.~\cite{Kie08}, and 
its main characteristics are briefly summarized in the following subsection.

\subsection{The HH Method for Bound States}
\label{subsec:bs}

The nuclear wave function for the three-body system 
with total angular momentum $J,J_z$ can be written as 
\begin{equation}
|\Psi^{JJ_z}\rangle=\sum_\mu c_\mu |\Psi^{JJ_z}_\mu\rangle \ ,
\label{eq:psi}
\end{equation}
where $|\Psi^{JJ_z}_\mu\rangle$ is a suitable complete set of 
states, and $\mu$ is an index denoting the set of quantum numbers 
necessary to completely specify the basis elements. 

The coefficients of the expansion can be calculated using the 
Rayleigh-Ritz variational principle, which states that
\begin{equation}
  \langle\delta_c \Psi^{JJ_z}\,|\,H-E\,|\Psi^{JJ_z}\rangle
   =0 \ ,
   \label{eq:rrvar}
\end{equation}
where $\delta_c \Psi^{JJ_z}$ indicates the variation of 
$\Psi^{JJ_z}$ for arbitrary infinitesimal 
changes of the linear coefficients $c_\mu$. 
The problem of determining $c_\mu$ and the energy $E$ 
is then reduced to a generalized eigenvalue problem, 
\begin{equation}
  \sum_{ \mu'}\,\langle\Psi^{JJ_z}_\mu\,|\,H-E\,|\, \Psi^{JJ_z}_{\mu'}\,\rangle \,c_{\mu'}=0
  \ .
  \label{eq:gepb}
\end{equation}
The main difficulty of the method is to compute the 
matrix elements of the Hamiltonian $H$ with respect to the basis states
$|\Psi^{JJ_z}_\mu\rangle$. Usually $H$ is given as a sum of terms (kinetic energy,
two-body potential, etc.). The calculation of the matrix elements of
some parts of $H$ can be more conveniently performed in coordinate 
space, while for other parts it could be easier to work in momentum
space. Therefore, it is important that the basis states 
$|\Psi^{JJ_z}_\mu\rangle$ have simple expressions in both spaces. The
HH functions indeed have such a property.

Let us first consider the expression of the HH functions in coordinate
space. The internal dynamics of a system of three nucleons of identical mass
$m$ is conveniently described in terms of the 
Jacobi vectors $\bfx_{1p},\bfx_{2p}$,
constructed from a given particle permutation denoted with $p$, which 
specifies the particle order 
$i,j,k$, and given by 
\begin{eqnarray}
  \bfx_{2p}&=&\frac{1}{\sqrt{2}}(\bfr_j-\bfr_i) \ , \nonumber \\
  \bfx_{1p}&=&\sqrt{\frac{2}{3}}(\bfr_k-\frac{1}{2}(\bfr_i+\bfr_j)) \ . 
  \label{eq:jacc3}
\end{eqnarray}
Here $p=1$ corresponds to the order 1,2,3.
It is convenient to replace the moduli of $\bfx_{2p}$ and 
$\bfx_{1p}$ with the so-called hyperradius and 
hyperangle, defined as~\cite{Fab83}
\begin{eqnarray}
  \rho&=&\sqrt{\bfx_{1p}^2+\bfx_{2p}^2} \ , \label{eq:rho} \\
  \tan{\phi_{p}}&=&\frac{x_{1p}}{x_{2p}} \ .  \label{eq:hypera}
\end{eqnarray}
Note that $\rho$ does not depend on
the particle permutation $p$. 
The complete set of hyperspherical coordinates is then 
given by $\{\rho,\Omega^{(\rho)}_p\}$, with 
\begin{equation} 
  \Omega^{(\rho)}_p=[{\hat{\bfx}}_{1p},{\hat{\bfx}}_{2p};\phi_{p}] \ , 
  \label{eq:omegar}
\end{equation}
and the suffix $(\rho)$ recalls the use of the coordinate space.  

The expansion states $|\Psi^{JJ_z}_\mu\rangle$ of 
Eq.~(\ref{eq:psi}) are then given by 
\begin{equation}
  |\,\Psi^{JJ_z(\rho)}_\mu\,\rangle
  = f_l(\rho) {\cal Y}_{ \{G\} }(\Omega^{(\rho)})\ ,
  \label{eq:rexp}
\end{equation}
where $f_l(\rho)$ for $l=1,\ldots\,M$ is a complete set of hyperradial 
functions, chosen of the form
\begin{equation}
 f_l(\rho)=\gamma^{3} \sqrt{\frac{l!}{(l+5)!}}\,\,\, 
 L^{(5)}_l(\gamma\rho)\,\,{\rm e}^{-\frac{\gamma}{2}\rho} \ .
 \label{eq:fllag}
\end{equation}
Here $L^{(5)}_l(\gamma\rho)$ are Laguerre polynomials, 
and the non-linear parameter $\gamma$ is 
variationally optimized. As an example, for the N3LO-Idaho potential, 
it can be chosen in the interval 6--8 fm$^{-1}$.

The functions ${\cal Y}_{ \{G\} }(\Omega^{(\rho)})$ are written 
as 
\begin{equation}
  {\cal Y}_{ \{G\} }(\Omega^{(\rho)})= \sum_{p=1}^{3}\bigg[ 
  Y^{LL_z}_{ [G] }(\Omega^{(\rho)}_p) 
  \otimes [S_2\otimes \frac{1}{2}]_{S S_z} \bigg]_{J J_z}\, 
  [T_2\otimes\frac{1}{2}]_{T T_z}   \ , \label{eq:hha3}
\end{equation}
where the sum is performed over the three even permutations.  
The spins (isospins) of particle $i$ and $j$ are coupled to 
$S_2$ ($T_2$), which is itself coupled to the spin (isospin) 
of the third particle to give 
the state with total spin $S$ (isospin $T,T_z$).
The total orbital angular momentum $L$ and the total 
spin $S$ are coupled to the total angular momentum 
$J,J_z$.  
The functions $Y^{LL_z}_{[G]}(\Omega^{(\rho)}_p)$, 
having  a definite value of 
$L,L_z$, are the HH functions, and 
are written as~\cite{Kie97a}:
\begin{equation}
  Y^{LL_z}_{ [G] }(\Omega^{(\rho)}_p) =
  \biggl[Y_{\ell_2}({\hat{\bfx}}_{2p}) \otimes
   Y_{\ell_1}({\hat{\bfx}}_{1p}) \biggr]_{LL_z}
  N_{[G] }\,(\cos\phi_p)^{\ell_2}(\sin\phi_p)^{\ell_1}\,
  P_{n}^{\ell_1+\frac{1}{2},\ell_2+\frac{1}{2}}(\cos 2\phi_p) \ .
\label{eq:hh3}
\end{equation}
Here $Y_{\ell_1}({\hat{\bfx}}_{1p})$ and $Y_{\ell_2}({\hat{\bfx}}_{2p})$ are 
spherical harmonics, $N_{[G]}$ is a normalization factor and 
$P_{n}^{\ell_1+\frac{1}{2},\ell_2+\frac{1}{2}}(\cos 2\phi_p)$ is a Jacobi 
polynomial, $n$ being the degree of the polynomial.
The grand angular quantum number $G$ is 
defined as $G=2n+\ell_1+\ell_2$.
The notations $[G]$ and $\{G\}$ of Eqs.~(\ref{eq:hh3}) 
and~(\ref{eq:hha3}) stand for $[\ell_1,\ell_2;n]$ and 
$\{\ell_1,\ell_2,L,S_2,T_2,S,T;n\}$, respectively, 
and $\mu$ of Eq.~(\ref{eq:rexp}) is $\mu=\{G\},l$. Note that 
each set of quantum numbers $\{\ell_1,\ell_2,L,S_2,T_2,S,T\}$ 
is called ``channel'', and the antisymmetrization of 
${\cal Y}_{ \{G\} }(\Omega^{(\rho)})$ requires $\ell_2+S_2+T_2$ 
to be odd. In addition, $\ell_1+\ell_2$ must be even (odd) 
for positive (negative) parity.
To be noticed that after the sum on the permutation in Eq.~(\ref{eq:hha3}), 
some states inside the sub-space spanned by $G$ are linearly dependent.
These states have been identified and removed from the 
expansion~\cite{Kie08,Kie97a}.

In this work, we have considered modern two-body potential models 
which act on specific spin and angular momentum states of the 
two-body system. Due to the presence of the sum over the permutations in
the expression for ${\cal Y}_{ \{G\} }$, 
a given particle pair
is not in a definite angular and spin state. However,
the HH functions with the grand angular quantum number $G$ constructed in
terms of a given set of Jacobi vectors
$\bfx_{1p},\bfx_{2p}$, defined starting from the particle 
order $i,j,k$, can be always expressed in terms of
the HH functions constructed, for instance, in terms of
$\bfx_{1 (p=1)},\bfx_{2 (p=1)}$.  
In fact, the following relation holds
\begin{equation}
  Y^{LL_z}_{[\ell_1,\ell_2;n]}(\Omega^{(\rho)}_p) =
   \sum_{\ell_1',\ell_2',n'}
    a^{(p),L}_{\ell_1,\ell_2,n;\,\ell_1',\ell_2',n'}
    Y^{LL_z}_{[\ell_1',\ell_2';n']}(\Omega^{(\rho)}_{(p=1)})\ ,
   \label{eq:rr3}
\end{equation}
where the sum is
restricted to the values $\ell_1'$, $\ell_2'$, and $n'$
such that $\ell_1'+\ell_2'+2n'=G$. 
The coefficients $ a^{(p),L}_{\ell_1,\ell_2,n;\,\ell_1',\ell_2',n'}$
relating the two sets of HH functions are known as the Raynal-Revai
coefficients~\cite{RR70}, and could be computed rather easily 
using the orthonormality property of the HH functions, namely 
\begin{equation}
  a^{(p),L}_{\ell_1,\ell_2,n;\,\ell_1',\ell_2',n'}=
  \int d\Omega^{(\rho)}_{(p=1)} \left(
   Y^{LL_z}_{[\ell_1,\ell_2;n]}(\Omega^{(\rho)}_{(p=1)})\right)^*
   Y^{LL_z}_{[\ell_1',\ell_2';n']}(\Omega^{(\rho)}_p)\ .
   \label{eq:rr3b}
\end{equation}
Also the spin-isospin states can be recoupled to obtain states where the
spin and isospin quantum numbers are coupled in a given order of the particles. 
The result is that the antisymmetric functions 
${\cal Y}_{ \{G\} }$ can be expressed as a
superposition of functions constructed in terms of a given order of particles
$i,j,k$, each one having the pair $i$,$j$ in a definite spin and
angular momentum state. When the two-body potential acts on the pair of
particles $i$,$j$, the effect of the projection is easily taken into
account. 

The expansion states of Eq.~(\ref{eq:psi}) 
in momentum space can be obtained as follows.
Let $\hbar\bfk_{1p},\hbar\bfk_{2p}$ be the conjugate Jacobi momenta 
of the Jacobi vectors, given by 
\begin{eqnarray}
  \hbar\bfk_{2p}&=&\frac{1}{\sqrt{2}}(\bfp_j-\bfp_i) \ , \nonumber \\
  \hbar\bfk_{1p}&=&\sqrt{\frac{2}{3}}(\bfp_k-\frac{1}{2}(\bfp_i+\bfp_j)) \ , 
  \label{eq:jacm3}
\end{eqnarray}
$\bfp_i$ being the momentum of the $i$-th particle. 
We then define a hypermomentum $Q$ and a set of 
angular-hyperangular variables as 
\begin{eqnarray}
  Q&=&\sqrt{\bfk_{1p}^2+\bfk_{2p}^2} \ ,   \nonumber \\
  \Omega^{(Q)}_p&=&[{\hat{\bfk}}_{2p},{\hat{\bfk}}_{1p};\varphi_{p}] \ , 
  \label{eq:hyperq}
\end{eqnarray}
where
\begin{equation}
  \tan{\varphi_{p}}=\frac{k_{1p}}{k_{2p}} \ .
  \label{eq:hyperaq}
\end{equation}
Then, the momentum-space version of the wave function 
given in Eq.~(\ref{eq:rexp}) is
\begin{equation}
  |\,\Psi^{JJ_z(Q)}_\mu\,\rangle=
   g_{ G,l }(Q) {\cal Y}_{ \{G\} }(\Omega^{(Q)}) \ , 
  \label{eq:qexp}
\end{equation}
where ${\cal Y}_{ \{G\} }(\Omega^{(Q)})$ is the same as 
${\cal Y}_{ \{G\} }(\Omega^{(\rho)})$ of Eq.~(\ref{eq:hha3})
with $\bfx_{ip}\rightarrow\bfk_{ip}$, 
and 
\begin{equation}
   g_{G,l}(Q)=(-i)^G\,\int_0^\infty d\rho\,
   \frac{\rho^{3}}{Q^{2}}\,
   J_{G+2}(Q\rho)\, f_{l}(\rho) \ .
\label{eq:vg}
\end{equation}
With the adopted form of $f_l(\rho)$ given in Eq.~(\ref{eq:fllag}), 
the corresponding functions $g_{G,l}(Q)$ can be easily calculated, 
and they are explicitly given in Ref.~\cite{Viv06}.

\subsection{The HH Method for Scattering States Below Deuteron 
Breakup Threshold}
\label{subsec:ss}

We consider here the extension of the 
HH technique to describe $N-d$ scattering states 
below deuteron breakup threshold, when both local 
and non-local interaction models are considered.

Following Ref.~\cite{Kie94}, 
the wave function $\Psi_{N-d}^{L S J J_z}$ describing the $N-d$
scattering state with incoming orbital angular momentum $L$ and channel spin
$S$, parity $\pi=(-)^L$, 
and total angular momentum $J, J_z$,
can be written as 
\begin{equation}
    \Psi_{N-d}^{LSJJ_z}=\Psi_C^{LSJJ_z}+\Psi_A^{LSJJ_z} \ ,
    \label{eq:psica}
\end{equation}
where $\Psi_C^{LSJJ_z}$ describes the system in the region where the particles
are close to each other and their mutual interactions are strong, 
while $\Psi_A^{LSJJ_z}$ describes the relative motion between the nucleon $N$
and the deuteron in the asymptotic region, where the $N-d$ nuclear 
interaction is negligible. The function $\Psi_C^{LSJJ_z}$, which has to 
vanish in the limit of large intercluster
separations, can be expanded on the HH basis as it has been done 
in the case of bound states. Therefore, 
applying Eq.~(\ref{eq:psi}), 
the function $\Psi_C^{LSJJ_z}$ can be casted in the form 
\begin{equation}
  |\Psi^{LSJJ_z}_C\rangle=\sum_{\mu}\, c_\mu\,
  |\Psi^{JJ_z}_\mu \rangle \ ,
  \label{eq:psis}
\end{equation}
where $|\Psi^{JJ_z}_\mu\rangle$ is defined in Eqs.~(\ref{eq:rexp}) 
and~(\ref{eq:qexp}) in coordinate- and momentum-space, respectively.

The function $\Psi_A^{LSJJ_z}$ is the appropriate 
asymptotic solution of the relative $N-d$ Schr\"odinger equation. 
It is written as a linear combination of the following functions, 
\begin{equation}
  \Omega_{LSJJ_z}^{\lambda}=\sum_{p=1}^3\Omega_{LSJJ_z}^{\lambda}(p) \ ,
  \label{eq:psiomp} 
\end{equation}
where the sum over $p$ has to be done over the three even permutations 
necessary to antisymmetrize the functions $\Omega_{LSJJ_z}^{\lambda}$, and
\begin{eqnarray}
  \Omega_{LSJJ_z}^{\lambda}(p)&=& \sum_{l=0,2}w_l(x_{2p})\,R^\lambda_L(y_p)
\Bigl\{\Bigl[ [Y_l(\hat{\bfx}_{2p})\otimes S_2]_1\otimes \frac{1}{2}\Bigr]_S
\otimes Y_{L}(\hat{\bfy}_p) \Bigr \}_{JJ_z} 
\nonumber \\
&\times&[ T_2\otimes \frac{1}{2} ]_{TT_z} \ .
  \label{eq:psiom}
\end{eqnarray}
Here the spin and isospin quantum numbers of particles $i$ and $j$ 
have been coupled to $S_2$ and $T_2$, with $S_2=1$, $T_2=0$ for the 
deuteron,  
$w_l(x_{2p})$ is the deuteron wave function component in the waves $l=0,2$, 
${\bfy}_p$ is the distance 
between $N$ and the center of mass of the deuteron, i.e. 
$\bfy_p=\sqrt{\frac{3}{2}}\bfx_{1p}$, 
$Y_l(\hat{\bfx}_{2p})$ and $Y_{L}(\hat{\bfy}_p)$ are the standard spherical 
harmonic functions, 
and the functions $R^\lambda_L(y_p)$ are the regular ($\lambda\equiv R$)
and irregular ($\lambda\equiv I$) radial solutions of the relative two-body 
$N-d$ Schr\"odinger equation without the nuclear interaction. 
These regular and irregular functions, denoted as 
${\cal F}_L(y_p)$ and ${\cal G}_L(y_p)$ respectively, have the form
\begin{eqnarray}
R^R_L(y_p)&\equiv&{\cal F}_L(y_p)=\frac{1}{(2L+1)!!q^L C_L(\eta)}\,{F_L(\eta,\xi_p)\over \xi_p} \ , 
\nonumber \\
R^I_L(y_p)&\equiv&{\cal G}_L(y_p)=(2L+1)!! q^{L+1}C_L(\eta)f_R(y_p){G_L(\eta,\xi_p)\over \xi_p}
\ ,
\label{eq:risol}
\end{eqnarray}
where $q$ is the
modulus of the $N-d$ relative momentum 
(related to the total kinetic energy in the center of mass system by
$T_{cm}={q^2\over 2\mu}$, $\mu$ being the $N-d$ reduced mass), 
$\eta=2\mu e^2/q$ and $\xi_p=qy_p$ are the usual Coulomb parameters, 
and the regular (irregular) Coulomb function $F_L(\eta,\xi_p)$ 
($G_L(\eta,\xi_p)$) and the 
factor $C_L(\eta)$ are defined in the standard 
way~\cite{chen:b}. The factor $(2L+1)!! q^L C_L(\eta)$
has been introduced so that ${\cal F}$ and ${\cal G}$
have a finite limit for $q\rightarrow 0$.
The function $f_R(y_p)=[1-\exp(-b y_p)]^{2L+1}$ 
has been introduced to regularize $G_L$ at small values of $y_p$. 
The trial parameter $b$ is
determined by requiring that $f_R(y_p)\rightarrow 1$ for 
large values of $y_p$,
thus not modifying the asymptotic behaviour of the 
scattering wave function. A value of $b=0.25$ fm$^{-1}$ 
has been found appropriate.
The non-Coulomb case of Eq.~(\ref{eq:risol}) is
obtained in the limit $e^2\rightarrow 0$. In this case, $F_L(\eta,\xi_p)/\xi_p$ and
$G_L(\eta,\xi_p)/\xi_p$ reduce to the regular and irregular Riccati-Bessel 
functions and
the factor $(2L+1)!!C_L(\eta)\rightarrow 1$ for $\eta\rightarrow 0$.

With the above definitions, $\Psi_A^{LSJJ_z}$  can be written in the form
\begin{equation}
  \Psi_A^{LSJJ_z}= \sum_{L^\prime S^\prime}
 \bigg[\delta_{L L^\prime} \delta_{S S^\prime} 
\Omega_{L^\prime S^\prime JJ_z}^R
  + {\cal R}^J_{LS,L^\prime S^\prime}(q)
     \Omega_{L^\prime S^\prime JJ_z}^I \bigg] \ ,
  \label{eq:psia}
\end{equation}
where the parameters ${\cal R}^J_{LS,L^\prime S^\prime}(q)$ give the
relative weight between the regular and irregular components 
of the wave function. They
are closely related to the reactance matrix (${\cal K}$-matrix)
elements, which can be written as
\begin{equation}
 {\cal K}^J_{LS,L^\prime S^\prime}(q)=
 (2L+1)!!(2L'+1)!!q^{L+L'+1}C_L(\eta)C_{L^\prime}(\eta)
 {\cal R}^J_{LS,L^\prime S^\prime}(q) \;\;\ .
\end{equation}
By definition of the ${\cal K}$-matrix, its eigenvalues are
$\tan\delta_{LSJ}$, $\delta_{LSJ}$ being the phase shifts.
The sum over $L^\prime$ and $S^\prime$ in Eq.~(\ref{eq:psia}) is over all 
values compatible with a given $J$ and parity $\pi$. In particular, the sum 
over $L^\prime$ is limited to include either even or odd values since
$(-1)^{L^\prime}=\pi$.

The matrix elements ${\cal R}^J_{LS,L^\prime S^\prime}(q)$ and 
the linear coefficients $c_\mu$ occurring in the expansion of $\Psi^{LSJJ_z}_C$ 
of Eq.~(\ref{eq:psis})
are determined applying the Kohn variational principle~\cite{kohn}, 
which states that the functional
\begin{eqnarray}
   [{\cal R}^J_{LS,L^\prime S^\prime}(q)]&=&
    {\cal R}^J_{LS,L^\prime S^\prime}(q)
     - \left \langle \Psi^{L^\prime S^\prime JJ_z }_{N-d} \left |
         {\cal L} \right |
        \Psi^{LSJJ_z}_{N-d}\right \rangle \ , \nonumber \\
{\cal L}&=&\frac{m}{2\sqrt{3}\hbar^2}(H-E) \ , \label{eq:kohn}
\end{eqnarray}
has to be stationary with respect to variations of the trial parameters 
in $\Psi^{LSJJ_z}_{N-d}$. 
Here $E$ is the total energy of the system, $m$ is the nucleon mass, 
and ${\cal L}$ is chosen so that 
\begin{equation}
   \langle \Omega^R_{LSJJ_z}| {\cal L} | \Omega^I_{LSJJ_z} \rangle
  -\langle \Omega^I_{LSJJ_z}| {\cal L} | \Omega^R_{LSJJ_z} \rangle =1 \ .
\end{equation}
As described in Ref.~\cite{Kie97}, 
using Eqs.~(\ref{eq:psis}) and~(\ref{eq:psia}), 
the variation of the diagonal functionals of Eq.~(\ref{eq:kohn}) with
respect to the linear parameters $c_\mu$ leads to the following 
system of linear inhomogeneous equations:
\begin{equation}
  \sum_{\mu'} \langle \Psi^{JJ_z}_\mu| {\cal L} |\Psi^{JJ_z}_{\mu'}\rangle c_{\mu'} = 
     -D^\lambda_{LSJJ_z}(\mu) \ .
  \label{eq:set1}
\end{equation}
Two different terms $D^\lambda$ corresponding to
$\lambda\equiv R,I$ are introduced and are defined as 
\begin{equation}
  D^\lambda_{LSJJ_z}(\mu)= \langle \Psi^{JJ_z}_\mu| {\cal L} |
\Omega^\lambda_{LSJJ_z}\rangle \ .
\label{eq:dlm}
\end{equation}
The matrix elements ${\cal R}^J_{LS,L'S'}(q)$ are obtained 
varying the diagonal functionals of Eq.~(\ref{eq:kohn}) with respect to them. 
This leads to the following set of algebraic equations
\begin{equation}
  \sum_{L'' S''} {\cal R}^J_{LS,L''S''}(q) X_{L'S',L''S''}= Y_{LS,L'S'} \ ,
\label{eq:set2}
\end{equation}
with the coefficients $X$ and $Y$ defined as
\begin{eqnarray}
X_{LS,L'S'}&= \langle
\Omega^I_{LSJJ_z}+\Psi^{LSJJ_z,I}_C| {\cal L} |\Omega^I_{L'S'JJ_z}\rangle \ ,
\nonumber \\
Y_{LS,L'S'}&=-\langle
\Omega^R_{LSJJ_z}+\Psi^{LSJJ_z,R}_C| {\cal L} |\Omega^I_{L'S'JJ_z}\rangle \ .
\label{eq:xy}
\end{eqnarray}
Here $\Psi^{LSJJ_z,\lambda}_C$ is the solution of the set of 
Eq.~(\ref{eq:set1}) with the corresponding $D^\lambda$ term. A 
second order estimate of ${\cal R}^J_{LS,L'S'}(q)$ is 
given by the quantities $[{\cal R}^J_{LS,L'S'}(q)]$, obtained by 
substituting in Eq.~(\ref{eq:kohn}) the
first order results. Such second-order calculation provides a symmetric 
reactance matrix. This condition is not {\it a priori} imposed, 
and therefore it is a useful test of the numerical accuracy.

In the particular case of $q=0$ (zero-energy scattering),
the scattering can occur only in the channel $L=0$ and the observables 
of interest are the scattering lengths. Within the 
present approach, they can be easily obtained from the relation
\begin{equation}
  ^{(2J+1)}a_{Nd}=-\lim_{q\rightarrow 0}{\cal R}^J_{0J,0J}(q)\ .
\label{eq:scleng}
\end{equation}

An alternative way to solve the scattering problem, used when 
$q\neq 0$, is to apply the complex Kohn variational principle 
to the ${\cal S}$-matrix, as in Ref.~\cite{Kie97}. In this way, the Kohn 
variational principle of Eq.~(\ref{eq:kohn}) becomes
\begin{equation}
   [{\cal S}^J_{LS,L^\prime S^\prime}(q)]=
    {\cal S}^J_{LS,L^\prime S^\prime}(q)
     + i \langle \Psi^{+,L^\prime S^\prime JJ_z}_{N-d} |
         {\cal S} |
        \Psi^{+,LSJJ_z}_{N-d} \rangle \ . \label{eq:cmplxkohn}
\end{equation}
Here
\begin{equation}
    \Psi_{N-d}^{+,LSJJ_z}=\Psi_C^{LSJJ_z}+\Psi_A^{+,LSJJ_z} \ ,
    \label{eq:cmplxpsica}
\end{equation}
with $\Psi_C^{LSJJ_z}$ given in Eq.~(\ref{eq:psis}) and
\begin{eqnarray}
\Psi_A^{+,LSJJ_z}&=&\sum_{p=1}^{3} \Omega^+_{LSJJ_z}(p) \nonumber \\
\Omega^+_{LSJJ_z}(p)&=&(\, i \tilde{\Omega}^R_{LSJJ_z}(p) -
\tilde{\Omega}^I_{LSJJ_z}(p)\,) \nonumber \\
&+&\sum_{L^\prime S^\prime}{\cal S}^J_{LS,L^\prime S^\prime}(q)
(\, i \tilde{\Omega}^R_{L^\prime S^\prime JJ_z}(p) +
\tilde{\Omega}^I_{L^\prime S^\prime JJ_z}(p)\, )
\ . \label{eq:cmplxpsia}
\end{eqnarray}
The functions $\tilde{\Omega}^\lambda_{LSJJ_z}(p)$ are the same as 
in Eq.~(\ref{eq:psiom}), with 
$R^R_L(y_p)=F_L(\eta,\xi_p)/ \xi_p$ and  
$R^I_L(y_p)=f_R(y_p) G_L(\eta,\xi_p)/ \xi_p$.
Note that, with the above definition, the reactance ${\cal K}$-matrix 
elements can be related to the ${\cal S}$-matrix elements as
\begin{equation}
{\cal K}^J_{LS,L^\prime S^\prime}(q)=(-i)
[{\cal S}^J_{LS,L^\prime S^\prime}(q) - \delta_{LL^\prime}\delta_{S S^\prime}]
\,
[{\cal S}^J_{LS,L^\prime S^\prime}(q) + \delta_{LL^\prime}\delta_{S S^\prime}]^{-1}
\ . \label{eq:ksmt}
\end{equation}

The calculation involving $\Psi_C^{LSJJ_z}$
has been performed with the HH expansion 
in coordinate- or in momentum-space, depending 
on what is more convenient, as it has been explained for the bound state 
in the previous subsection. Some difficulties arise for the 
calculation of the potential energy matrix elements which involve 
$\Omega^\lambda_{LSJJ_z}$, i.e. 
$\langle\Psi^{JJ_z}_\mu |V|\Omega^\lambda_{LSJJ_z}\rangle$ present in 
Eq.~(\ref{eq:dlm}), and 
$\langle\Omega^{\lambda '}_{L'S'JJ_z}+\Psi_C^{L'S'JJ_Z,\lambda'} |V|
\Omega^\lambda_{LSJJ_z}\rangle$ of Eq.~({\ref{eq:xy}), 
with $\lambda, \lambda '=R$, $I$. 
In the present work, we consider both two- and three-nucleon interactions, 
and therefore
\begin{equation}
V=\sum_{i<j}V_{ij}+\sum_{i<j<k}V_{ijk} \ .
\label{eq:pot}
\end{equation}
We first focus on the two-body contribution.
Due to the 
antisymmetry of the wave function,  the following relation holds
\begin{equation}
\langle\Phi|\sum_{i<j}V_{ij}|\Omega_{LSJJ_z}^{\lambda }(p)\rangle = 
3 \langle\Phi|V_{12}|\Omega_{LSJJ_z}^{\lambda }(p)\rangle \ , 
\label{eq:v12}
\end{equation}
where $|\Phi\rangle$ can be either $|\Psi^{JJ_z}_\mu\rangle$ 
of Eq.~(\ref{eq:psis}) or 
$\Omega_{L'S'JJ_z}^{\lambda '}(p')$ of Eq.~(\ref{eq:psiom}),
depending on which term among $D^\lambda$, $X$, and $Y$ is considered.
The potential $V_{12}$ acts on the particle pair 1,2 with
total angular momentum $j$, and orbital angular momentum and 
spin quantum numbers $\ell_{12}',s_{12}'$ (on the bra) and 
$\ell_{12},s_{12}$ (on the ket), and
can be written as
\begin{equation}
V_{12}=v_{12}^j(x_{2\,(p=1)}',x_{2\,(p=1)};
\ell_{12}',s_{12}',\ell_{12},s_{12})\ ,
\label{eq:v12r}
\end{equation}
in coordinate-space, and 
\begin{equation}
V_{12}=v_{12}^j(k_{2\,(p=1)}',k_{2\,(p=1)};
\ell_{12}',s_{12}',\ell_{12},s_{12})\ ,
\label{eq:v12p}
\end{equation}
in momentum-space, where $x_{2\,(p=1)}$ and $k_{2\,(p=1)}$ 
are the moduli of the vectors defined in Eqs.~(\ref{eq:jacc3}) 
and~(\ref{eq:jacm3}), respectively. 
When local potential models are considered, such as the AV18,
then 
\begin{equation}
v_{12}^j(x_{2\,(p=1)}',x_{2\,(p=1)};\ell_{12}',s_{12}',\ell_{12},s_{12})\rightarrow
v_{12}^j(x_{2\,(p=1)};\ell_{12}',s_{12}',\ell_{12},s_{12})
\delta(x_{2\,(p=1)}-x_{2\,(p=1)}')\ .
\label{eq:v12loc}
\end{equation}
The first difficulty that needs to be overcame arises from the fact that 
when the $V_{12}$ operator acts on $\Omega_{LSJJ_z}^\lambda(p\neq 1)$, 
the particle pair 12 does not have a well definite 
orbital and spin angular momenta. However, the following relation holds: 
\begin{eqnarray}
w_l(x_{2 (p\neq 1)})\,R^\lambda_L(y_{p\neq 1}) 
\Bigl[ Y_l({\hat{\bfx}}_{2 (p\neq 1)}) \otimes
Y_L({\hat{\bfy}}_{p\neq 1}) \Bigl]_{\Lambda,\Lambda_z} &=& 
\sum_{l',L'} F_{lL;l'L'}^{\lambda,p\neq 1}(x_{1\,(p=1)},x_{2\,(p=1)}) \nonumber \\
&\times&
\Bigl[ Y_{l'}({\hat{\bfx}}_{2 (p=1)}) \otimes
Y_{L'}({\hat{\bfy}}_{p=1}) \Bigl]_{\Lambda,\Lambda_z} \ . \label{eq:om12}
\end{eqnarray}
where $\Lambda,\Lambda_z$ are the total orbital angular momentum and 
its third component.
The functions $F_{lL;l'L'}^{\lambda,p\neq 1}(x_{1 (p=1)},x_{2 (p=1)})$
are given by
\begin{eqnarray}
F_{lL;l'L'}^{\lambda,p\neq 1}(x_{1 (p=1)},x_{2 (p=1)}) &=& 
\int d{\hat{\bfx}}_{2 (p=1)} d{\hat{\bfx}}_{1 (p=1)} 
\Bigl[ Y_{l'}^*({\hat{\bfx}}_{2 (p=1)}) \otimes 
Y_{L'}^*({\hat{\bfy}}_{p=1}) \Bigl]_{\Lambda,\Lambda_z} 
\nonumber \\
&\times&w_l(x_{2 (p\neq 1)})\,
R^\lambda_L(y_{p\neq 1}) 
\Bigl[ Y_l({\hat{\bfx}}_{2 (p\neq 1)}) \otimes Y_L({\hat{\bfy}}_{p\neq 1}) 
\Bigl]_{\Lambda,\Lambda_z}
\ . \label{eq:fll}
\end{eqnarray}
Once the functions $F_{lL;l'L'}^{\lambda,p\neq 1}(x_{1 (p=1)},x_{2 (p=1)})$ 
have been calculated 
and the spin-isospin states have been also properly recoupled, 
the effect of the projection operator 
in $V_{12}$ is easily taken into account. 

A second difficulty arises in the calculation of the potential 
matrix element, when non-local potentials expressed in 
momentum-space are used. On the contrary to the 
core part of the scattering wave function $\Psi^{LSJJ_z}_C$, which can be 
alternatively expressed in coordinate- or in momentum-space, 
the asymptotic states $\Omega_{LSJJ_z}^{\lambda}$ do 
not have an easy expression in momentum-space, and are more conveniently 
expressed and used in coordinate-space. 
This is especially true when the Coulomb 
interaction is considered, as for the $p-d$ case. 
Therefore, we have decided to 
perform the Fourier transform of the potential 
$v_{12}^j(k',k;\ell',s',\ell,s)$ 
to work only in coordinate space, namely
\begin{equation}
v_{12}^j(r',r;\ell',s',\ell,s) =
\frac{2}{\pi}\int k^2 dk\,\,k'^2 dk'\,j_{\ell'}(k'r')\,
v_{12}^j(k',k;\ell',s',\ell,s)\,j_{\ell}(kr)
\ , \label{eq:ftv}
\end{equation}
where $j_{\ell}(kr)$ and $j_{\ell'}(k'r')$ 
are the standard spherical Bessel functions. 
The integrations over $k$ and $k'$, which run from 0 to $\infty$, are 
easily performed when the potential model considered does not have a 
high-momentum tail, but goes rapidly to zero at rather low values of 
$k$ and $k'$. This is true for the N3LO-Idaho
and $V_{low\!-\!k}$ potential models, but not for the 
CDBonn. Since the main goal of the present work 
is to perform a first test of the applicability of the HH method to the 
$A=3$ scattering problem using non-local realistic interactions, only the 
N3LO-Idaho and $V_{low\!-\!k}$ two-body potentials have been considered.

Some remarks about the calculation of the 
three-body contribution to the 
potential energy operator of Eq.~(\ref{eq:pot}) are in order.
The TNIs considered in the present work 
are the Urbana IX~\cite{Pud95} (UIX) and the N2LO~\cite{Nav07} potentials. 
The first one is used in conjunction with both the AV18 and N3LO-Idaho 
two-nucleon interactions. In the second case, the parameter in 
front of the spin-isospin independent part of the UIX TNI has been 
rescaled by a factor of 0.384 to fit the triton binding 
energy~\cite{Mar08} (UIXp). 
The N2LO TNI has been used only in conjunction with the 
N3LO-Idaho potential model. All these TNIs are local potentials, 
and have a well defined operatorial structure. Therefore, 
the projection procedure of Eqs.~(\ref{eq:om12}) and~(\ref{eq:fll}) 
is not needed and the present approach
follows the footsteps of the PHH technique~\cite{Kie94,Kie96}.

\section{Results}
\label{sec:res}

In this section we present our results for $n-d$ and $p-d$ scattering 
observables at center-of-mass energies below deuteron breakup 
threshold. The interaction models which have been used are 
the AV18 and the N3LO-Idaho
two-nucleon, and the 
AV18/UIX, N3LO-Idaho/UIXp and the N3LO-Idaho/N2LO two- and three-nucleon 
interactions. Note that the AV18 and AV18/UIX results are the same 
as those ones first obtained in Ref.~\cite{Kie95}, using the PHH expansion.
We have considered also the 
$V_{low\!-\!k}$ model, obtained from the AV18 two-nucleon 
interaction with a cutoff parameter $\Lambda$ equal to 2.2 fm$^{-1}$. 
The cutoff parameter has been chosen so that
the triton binding energy is
8.477 MeV, when the complete electromagnetic interaction is used, 
including neutron charge distribution and MM interaction 
effects. On the other hand, when no electromagnetic effects are considered, 
the triton binding energy has been found to be 
8.519 MeV. In the scattering problem, only the point Coulomb interaction 
has been considered, except when differently indicated.

Before presenting the results for the considered low-energy $N-d$ observables,
we discuss the pattern of convergence for some representative quantities,
i.e.\ the $n-d$ doublet zero-energy scattering length $^2a_{nd}$
and the $p-d$ $J^\pi=1/2^+,1/2^-$ phase shifts and mixing 
angles at center-of-mass energy 
$E_{cm}=2.0$ MeV, calculated with the N3LO-Idaho two-nucleon interaction 
model. The angular momentum-spin-isospin channels considered 
for $J^\pi=1/2^+$ and $1/2^-$ are given in Tables~\ref{tab:ch+} 
and~\ref{tab:ch-}, respectively. The notation is the same as 
in Eq.~(\ref{eq:hh3}). To be noticed that the scattering channels in the 
case of $J^\pi=1/2^-$ are ordered for increasing values of 
$\ell_{1}+\ell_{2}$. This is true also for all the channels here 
considered,
except those for $J^\pi=1/2^+$ (see Table~\ref{tab:ch+}), where the ordering 
respects an ``historical choice'', first done in the case of the three-nucleon 
bound state in Ref.~\cite{Kam89}.
\begin{table}
\caption{\label{tab:ch+}
Angular momentum, spin and isospin quantum numbers for the first 23
channels considered in the expansion of the $J^\pi =1/2^+$ 
core wave function.}
\begin{center}
\begin{tabular}{cccccccc}
\hline
   $\alpha$ &$\ell_{1\alpha}$&$\ell_{2\alpha}$&$L_\alpha$&
      $S_{2\alpha}$ & $T_{2\alpha}$& $S_\alpha$ & $T_\alpha$ \\
\hline
 1 & 0 & 0 & 0 & 1 & 0 & 1/2 & 1/2 \\
 2 & 0 & 0 & 0 & 0 & 1 & 1/2 & 1/2 \\
 3 & 0 & 2 & 2 & 1 & 0 & 3/2 & 1/2 \\
 4 & 2 & 0 & 2 & 1 & 0 & 3/2 & 1/2 \\
 5 & 2 & 2 & 0 & 1 & 0 & 1/2 & 1/2 \\
 6 & 2 & 2 & 2 & 1 & 0 & 3/2 & 1/2 \\
 7 & 2 & 2 & 1 & 1 & 0 & 1/2 & 1/2 \\
 8 & 2 & 2 & 1 & 1 & 0 & 3/2 & 1/2 \\
 9 & 1 & 1 & 0 & 1 & 1 & 1/2 & 1/2 \\
10 & 1 & 1 & 1 & 1 & 1 & 1/2 & 1/2 \\
11 & 1 & 1 & 1 & 1 & 1 & 3/2 & 1/2 \\
12 & 1 & 1 & 2 & 1 & 1 & 3/2 & 1/2 \\
13 & 1 & 1 & 0 & 0 & 0 & 1/2 & 1/2 \\
14 & 1 & 1 & 1 & 0 & 0 & 1/2 & 1/2 \\
15 & 2 & 2 & 0 & 0 & 1 & 1/2 & 1/2 \\
16 & 2 & 2 & 1 & 0 & 1 & 1/2 & 1/2 \\
17 & 3 & 1 & 2 & 1 & 1 & 3/2 & 1/2 \\
18 & 1 & 3 & 2 & 1 & 1 & 3/2 & 1/2 \\
19 & 0 & 0 & 0 & 0 & 1 & 1/2 & 3/2 \\
20 & 1 & 1 & 0 & 1 & 1 & 1/2 & 3/2 \\
21 & 1 & 1 & 1 & 1 & 1 & 1/2 & 3/2 \\
22 & 1 & 1 & 1 & 1 & 1 & 3/2 & 3/2 \\
23 & 1 & 1 & 2 & 1 & 1 & 3/2 & 3/2 \\
\hline
\end{tabular}
\end{center}
\end{table}
\begin{table}
\caption{\label{tab:ch-}
Same as Table~\protect\ref{tab:ch+}, but for the first 25
channels considered in the expansion of the $J^\pi =1/2^-$ 
core wave function.}
\begin{center}
\begin{tabular}{cccccccc}
\hline
   $\alpha$ &$\ell_{1\alpha}$&$\ell_{2\alpha}$&$L_\alpha$&
      $S_{2\alpha}$ & $T_{2\alpha}$& $S_\alpha$ & $T_\alpha$ \\
\hline
 1 & 1 & 0 & 1 & 1 & 0 & 1/2 & 1/2 \\
 2 & 1 & 0 & 1 & 0 & 1 & 1/2 & 1/2 \\
 3 & 1 & 0 & 1 & 1 & 0 & 3/2 & 1/2 \\
 4 & 0 & 1 & 1 & 1 & 1 & 1/2 & 1/2 \\
 5 & 0 & 1 & 1 & 0 & 0 & 1/2 & 1/2 \\
 6 & 0 & 1 & 1 & 1 & 1 & 3/2 & 1/2 \\
 7 & 2 & 1 & 1 & 1 & 1 & 1/2 & 1/2 \\
 8 & 2 & 1 & 1 & 0 & 0 & 1/2 & 1/2 \\
 9 & 2 & 1 & 1 & 1 & 1 & 3/2 & 1/2 \\
10 & 2 & 1 & 2 & 1 & 1 & 3/2 & 1/2 \\
11 & 1 & 2 & 1 & 1 & 0 & 1/2 & 1/2 \\
12 & 1 & 2 & 1 & 0 & 1 & 1/2 & 1/2 \\
13 & 1 & 2 & 1 & 1 & 0 & 3/2 & 1/2 \\
14 & 1 & 2 & 2 & 1 & 0 & 3/2 & 1/2 \\
15 & 3 & 2 & 1 & 1 & 0 & 1/2 & 1/2 \\
16 & 3 & 2 & 1 & 0 & 1 & 1/2 & 1/2 \\
17 & 3 & 2 & 1 & 1 & 0 & 3/2 & 1/2 \\
18 & 3 & 2 & 2 & 1 & 0 & 3/2 & 1/2 \\
19 & 1 & 0 & 1 & 0 & 1 & 1/2 & 3/2 \\
20 & 0 & 1 & 1 & 1 & 1 & 1/2 & 3/2 \\
21 & 0 & 1 & 1 & 1 & 1 & 3/2 & 3/2 \\
22 & 2 & 1 & 1 & 1 & 1 & 1/2 & 3/2 \\
23 & 2 & 1 & 1 & 1 & 1 & 3/2 & 3/2 \\
24 & 2 & 1 & 2 & 1 & 1 & 3/2 & 3/2 \\
25 & 1 & 2 & 1 & 0 & 1 & 1/2 & 3/2 \\
\hline
\end{tabular}
\end{center}
\end{table}

In Table~\ref{tab:conv1} we present 
the results for $^2a_{nd}$ and 
$p-d$ $J^\pi=1/2^+,1/2^-$ phase shifts and mixing 
angles $(\delta_{LSJ},\epsilon)$ at 
$E_{cm}=2.0$ MeV, for increasing 
values of the Laguerre polynomials $M$ in the hyperradial 
functions (see Eqs.~(\ref{eq:fllag}) and~(\ref{eq:vg})). 
All the 23 (25) angular momentum-spin-isospin channels 
of Table~\ref{tab:ch+} (\ref{tab:ch-}) are considered 
for $J^\pi=1/2^+$ ($1/2^-$), and HH functions up to 
grand angular momentum 
$G=20$ (21) for all the channels have been included.
From inspection of the table, we can conclude that the use of 
$M=28$ is enough to reach an accuracy of at least 0.002 fm 
for the scattering length and four 
significant digits for the phase shifts and mixing angles. 
In fact, for other $p-d$ scattering channels at some of the considered 
values of $E_{cm}$, even $M=24$ and $M=20$ has been found enough 
to reach the same degree of accuracy.

\begin{table}
\caption{\label{tab:conv1}
$n-d$ doublet scattering length $^2a_{nd}$ in fm 
and $p-d$ $J^\pi=1/2^+,1/2^-$ phase shifts  $\delta_{LSJ}$ and mixing 
angles $\epsilon$ at $E_{cm}=2.0$ MeV, 
calculated with the HH technique using the N3LO-Idaho interaction model,
for increasing values of the Laguerre polynomials $M$. All the channels 
of Tables~\protect\ref{tab:ch+} and~\protect\ref{tab:ch-} are included 
with grand angular momentum for all the channels set equal to 
20 for $J^\pi=1/2^+$ and 21 for $J^\pi=1/2^-$.}
\begin{center}
\begin{tabular}{cccccccc}
\hline
          & $M=4$  & $M=8$  & $M=12$  & $M=16$  & $M=20$  & $M=24$  & $M=28$  \\
\hline
$^2a_{nd}$ & 3.029  & 1.630  & 1.329   & 1.259   & 1.240   & 1.234  &  1.233 \\
\hline
$\delta_{0,\frac{1}{2},\frac{1}{2}}$ &
           -3.611  & -3.583 & -3.572 & -3.570 & -3.570 & -3.569 & -3.569 \\
$\delta_{2,\frac{3}{2},\frac{1}{2}}$ & 
           -43.28 &  -34.69 &  -32.41 & -31.96 & -31.82 & -31.78 & -31.77 \\
$\epsilon_{{\frac{1}{2}}^+}$ &
           0.525 &  0.975 & 1.150 & 1.189 & 1.201 & 1.205 & 1.206 \\
\hline
$\delta_{1,\frac{1}{2},\frac{1}{2}}$ &
           -8.270 & -7.756 & -7.608 & -7.581 & -7.576 & -7.575 & -7.575 \\
$\delta_{1,\frac{3}{2},\frac{1}{2}}$ & 
           20.82 & 21.73 & 21.97 & 22.00 & 22.00 & 22.01 & 22.01 \\
$\epsilon_{{\frac{1}{2}}^-}$ &
           4.947 & 5.542 & 5.628 & 5.641 & 5.643 & 5.644 & 5.644 \\
\hline
\end{tabular}
\end{center}
\end{table}

To study the convergence on the HH expansion, as it has been 
done in Ref.~\cite{Viv05}, 
we have separated the HH functions into classes having particular 
properties and we have taken into account the fact that the 
convergence rates of the different classes are rather different. 
For instance, we expect that 
the contribution of the HH functions with lower values of 
$\ell_{12,\alpha}=\ell_{1,\alpha}+\ell_{2,\alpha}$ to be the most important. 
Therefore, for all the $J^\pi$ scattering states, except $J^\pi = 1/2^+$, 
the different classes are 
classified with increasing value of $\ell_{12,\alpha}$, up to 
$\ell_{12,\alpha}\leq 6$,
and among those ones with the same $\ell_{12,\alpha}$, we have included 
first the contributions of the HH functions with lower 
$\ell_{2,\alpha}$. 
Finally, the $T_\alpha=3/2$ states are considered. 
With these criteria, in the $J^\pi = 1/2^-$ case, 
the channels have been classified in 6
classes, including channels 1--3, 4--6, 7--10, 11--14, 15--18, and 19--25 
of Table~\ref{tab:ch-}, respectively.
In the case of $J^\pi = 1/2^+$, the classification follows 
the footsteps of Ref.~\cite{Kam89}, 
and therefore the channels have been 
classified in 5 classes, including channels 1--3, 4--8, 9--12, 13--18, and 
19--23 of Table~\ref{tab:ch+}, respectively. 
We have then called $G_i$, for each class $i$, a number such 
that each state of class $i$ has the grand angular momentum $G\leq G_i$, 
and we have increased $G_i$ until we have reached convergence.
Then, keeping $G_i$ fixed at this value, we have included the states 
of the following class, and increased $G_{i+1}$ again until 
we have reached convergence.
The results for the zero-energy scattering length and the low-energy 
phase shifts and mixing angles obtained with this procedure are 
given in Table~\ref{tab:conv2+} for $J^\pi=1/2^+$ 
and~\ref{tab:conv2-} for $J^\pi=1/2^-$. Here, $M=28$ Laguerre polynomials 
in the expansion of the hyperradial function are included, and again
the N3LO-Idaho two-nucleon potential is used.

From the cases presented in the Tables, and as well as for all cases
taken in consideration, we can observe that
(i) the last classes of channels, corresponding to the $T_\alpha=3/2$ states,
give sizable contributions to the $p-d$ phase shifts and mixing angles, 
but negligible ones to the $n-d$ ones. 
(ii) The $T_\alpha=1/2$ channels with the largest values of $\ell_{12,\alpha}$ 
(fourth class for $J^\pi=1/2^+$ and fifth one for $J^\pi=1/2^-$) give negligible 
contributions. This implies that $\ell_{12,\alpha}\leq 6$ is enough 
to have accurate results.
(iii) The convergence with respect to the grand angular momentum for the first class
is the most problematic and it depends noticeably on the interaction. For
example, for $J^\pi=1/2^+$, when the 
non-local potential N3LO-Idaho is used, values of $G_1$ up to 80 have been
found to be necessary (see Table~\ref{tab:conv2+}). On the other hand, 
in the case of the local AV18, we have verified 
that within the HH expansion (i.e.\ without the correlation) 
$G_1=160$ is needed to reach the same degree of accuracy.
This is related to the fact that the AV18 potential is more repulsive at short
interparticle distances, and therefore the corresponding wave functions in
that region are more difficult to be constructed. In fact, when the
calculation is performed using  the $V_{low\!-\!k}$ potential model, which is
very soft at short interparticle distances, it is sufficient to set
$G_1=40$. A completely identical pattern of convergence is found for all other
$J^\pi$ waves. (iv) The convergence of the other classes is usually faster
than for the first class, as it is evident for the cases reported in
Tables~\ref{tab:conv2+} and ~\ref{tab:conv2-}. For the $J^\pi=1/2^+$ case,
we obtain convergence with just $G_{2,3,4,5}=20$. For
$J^\pi=1/2^-$, we have to consider fairly large values of $G$ only for the
fourth class (up to $G_4=51$), since the channels belonging to this class
(the channels 11-14 as reported in Table~\ref{tab:ch-}) are needed to
describe pairs in orbital angular momentum $\ell_2=2$. Namely, together with the
channels of the first class, they are needed to have a good descriptions of the
pairs in the deuteron waves. We have also found that the convergence rate of
these classes does not depend much on the 
non-local interaction model. For example, with the $V_{low\!-\!k}$ potential,
convergence is achieved with $G_{2,3,4,5}=20$ for $J^\pi=1/2^+$ and
$G_{2,4}=31$, $G_{3,5,6}=21$  for $J^\pi=1/2^-$.
However, note that for the AV18 potential model, we need to set 
$G_2=90$, $G_{3,4,5}=40$ for $J^\pi=1/2^+$ and
$G_2=61$, $G_{3,6}=41$, $G_4=91$, $G_5=21$ for $J^\pi=1/2^-$.
A similar pattern of convergence has been found 
for all the calculated quantities. From now on, all the results which 
will be presented have been obtained at convergence in the basis
expansion.

\begin{table}
\caption{\label{tab:conv2+} Convergence of the 
$n-d$ doublet scattering length $^2a_{nd}$ in fm
and $p-d$ $J^\pi=1/2^+$ phase shifts  $\delta_{LSJ}$ and mixing 
angles $\epsilon$ at $E_{cm}=2.0$ MeV, 
corresponding to the inclusion in the wave function of the 
different classes in which the HH basis has been divided. 
The N3LO-Idaho interaction model is used, and 
the number of Laguerre polynomials included is $M=28$. }
\begin{center}
\begin{tabular}{ccccccccc}
\hline
$G_1$ & $G_2$ & $G_3$ & $G_4$ & $G_5$ & 
$^2a_{nd}$ & 
$\delta_{0,\frac{1}{2},\frac{1}{2}}$ & 
$\delta_{2,\frac{3}{2},\frac{1}{2}}$ & 
$\epsilon_{{\frac{1}{2}}^+}$ \\
\hline
50 &    &    &    &    & 1.245 & -3.577 & -32.11 & 1.248 \\
60 &    &    &    &    & 1.243 & -3.577 & -32.09 & 1.248 \\
70 &    &    &    &    & 1.242 & -3.577 & -32.08 & 1.248 \\
80 &    &    &    &    & 1.242 & -3.577 & -32.08 & 1.248 \\
80 & 16 &    &    &    & 1.112 & -3.572 & -31.18 & 1.240 \\
80 & 20 &    &    &    & 1.112 & -3.572 & -31.17 & 1.238 \\
80 & 20 & 16 &    &    & 1.100 & -3.569 & -31.09 & 1.239 \\
80 & 20 & 20 &    &    & 1.100 & -3.569 & -31.09 & 1.239 \\
80 & 20 & 20 & 16 &    & 1.099 & -3.569 & -31.09 & 1.239 \\
80 & 20 & 20 & 20 &    & 1.099 & -3.569 & -31.09 & 1.239 \\
80 & 20 & 20 & 20 & 16 & 1.099 & -3.569 & -31.04 & 1.241 \\
80 & 20 & 20 & 20 & 20 & 1.099 & -3.569 & -31.04 & 1.241 \\
\hline
\end{tabular}
\end{center}
\end{table}
\begin{table}
\caption{\label{tab:conv2-} Same as Table~\protect\ref{tab:conv2+}
but for $p-d$ $J^\pi=1/2^-$ phase shifts  $\delta_{LSJ}$ and mixing 
angles $\epsilon$ at $E_{cm}=2.0$ MeV. }
\begin{center}
\begin{tabular}{ccccccccc}
\hline
$G_1$ & $G_2$ & $G_3$ & $G_4$ & $G_5$ & $G_6$ & 
$\delta_{1,\frac{1}{2},\frac{1}{2}}$ & 
$\delta_{1,\frac{3}{2},\frac{1}{2}}$ & 
$\epsilon_{{\frac{1}{2}}^-}$ \\
\hline
61 &    &    &    &    &    & -7.416 & 21.24 & 5.544 \\
71 &    &    &    &    &    & -7.413 & 21.25 & 5.545 \\
81 &    &    &    &    &    & -7.412 & 21.25 & 5.545 \\
91 &    &    &    &    &    & -7.411 & 21.25 & 5.545 \\
91 & 11 &    &    &    &    & -7.382 & 21.53 & 5.619 \\
91 & 21 &    &    &    &    & -7.380 & 21.55 & 5.622 \\
91 & 31 &    &    &    &    & -7.379 & 21.55 & 5.622 \\
91 & 31 & 15 &    &    &    & -7.367 & 21.77 & 5.704 \\
91 & 31 & 21 &    &    &    & -7.367 & 21.77 & 5.705 \\
91 & 31 & 21 & 31 &    &    & -7.372 & 22.02 & 5.799 \\
91 & 31 & 21 & 41 &    &    & -7.370 & 22.03 & 5.798 \\
91 & 31 & 21 & 51 &    &    & -7.369 & 22.04 & 5.798 \\
91 & 31 & 21 & 51 & 15 &    & -7.369 & 22.04 & 5.798 \\
91 & 31 & 21 & 51 & 21 &    & -7.369 & 22.04 & 5.798 \\
91 & 31 & 21 & 51 & 21 & 15 & -7.342 & 22.05 & 5.818 \\
91 & 31 & 21 & 51 & 21 & 21 & -7.340 & 22.05 & 5.819 \\
\hline
\end{tabular}
\end{center}
\end{table}

The results for the $n-d$ and $p-d$ doublet and quartet scattering lengths 
are given in Table~\ref{tab:sl} and are compared with the 
available experimental data~\cite{Dil71,Sch03}.
The results 
for the AV18 and AV18/UIX have been taken from Ref.~\cite{Kie08}.
Comparing the theoretical and experimental results for 
$^2a_{nd}$ and $^4a_{nd}$, we can conclude that $^4a_{nd}$ is 
very little model-dependent (as well as $^4a_{pd}$), and there is 
a satisfactory agreement between theory and experiment. On the contrary,
$^2a_{nd}$ is strongly model-dependent, and only 
the inclusion of the TNI brings the theoretical 
value close to the experimental one. However, 
some disagreement still remains, and the recent measurement
of Ref.~\cite{Sch03} is not well described by any of the
potential models considered. Though, the N3LO-Idaho/UIXp and 
N3LO-Idaho/N2LO models give slightly better results. 
Note that the AV18/UIX results obtained including also 
MM interaction effects are $^2a_{nd}=0.590$ fm and $^4a_{nd}=6.343$ fm.
Finally, the $V_{low\!-\!k}$ results are in remarkable disagreement with the 
experimental data, and a sizable difference 
from the AV18/UIX results is also observed. 
Therefore, even when the cutoff parameter of the 
$V_{low\!-\!k}$ interaction model is 
fixed to reproduce the triton binding energy, the doublet 
scattering length is not well reproduced. 
This observation seems to suggest that the $S$-wave sensitive 
scattering observables, like the scattering lengths, are not 
properly described by simply increasing the attraction, but a 
right balance 
between attraction and repulsion of the nuclear force has to be reached.
Such a balance
cannot be achieved with just one parameter, as the cutoff $\Lambda$ 
of the $V_{low\!-\!k}$ interaction.
Further analysis of these aspects is currently underway~\cite{Kie09}. 
\begin{table}
\caption{\label{tab:sl}
$n-d$ and $p-d$ doublet and quartet scattering lengths in fm
calculated with the HH technique using different Hamiltonian models.}
\begin{center}
\begin{tabular}{ccccc}
\hline
Interaction & $^2a_{nd}$ & $^4a_{nd}$ & $^2a_{pd}$ & $^4a_{pd}$ \\
\hline
AV18 & 1.275 & 6.325 & 
1.185 & 13.588 \\
AV18/UIX & 0.610 & 6.323 & 
-0.035 & 13.588 \\
N3LO-Idaho & 1.099 & 6.342 & 
0.876 & 13.646 \\
N3LO-Idaho/UIXp & 0.623 & 6.343 & 
-0.007 & 13.647 \\
N3LO-Idaho/N2LO & 0.675 & 6.342 & 
0.072 & 13.647 \\
$V_{low\!-\!k}$ & 0.572 & 6.321 & -0.001 & 13.571 \\
\hline
Exp.{~\protect{\cite{Dil71}}} & 0.65$\pm$0.04 & 6.35 $\pm$0.02 & & \\
Exp.{~\protect{\cite{Sch03}}} & 0.645$\pm$0.003$\pm$0.007 & -- & & \\
\hline
\end{tabular}
\end{center}
\end{table}
\begin{figure}[tbp]
\begin{center}
\includegraphics[width=\textwidth,height=0.5\textheight]{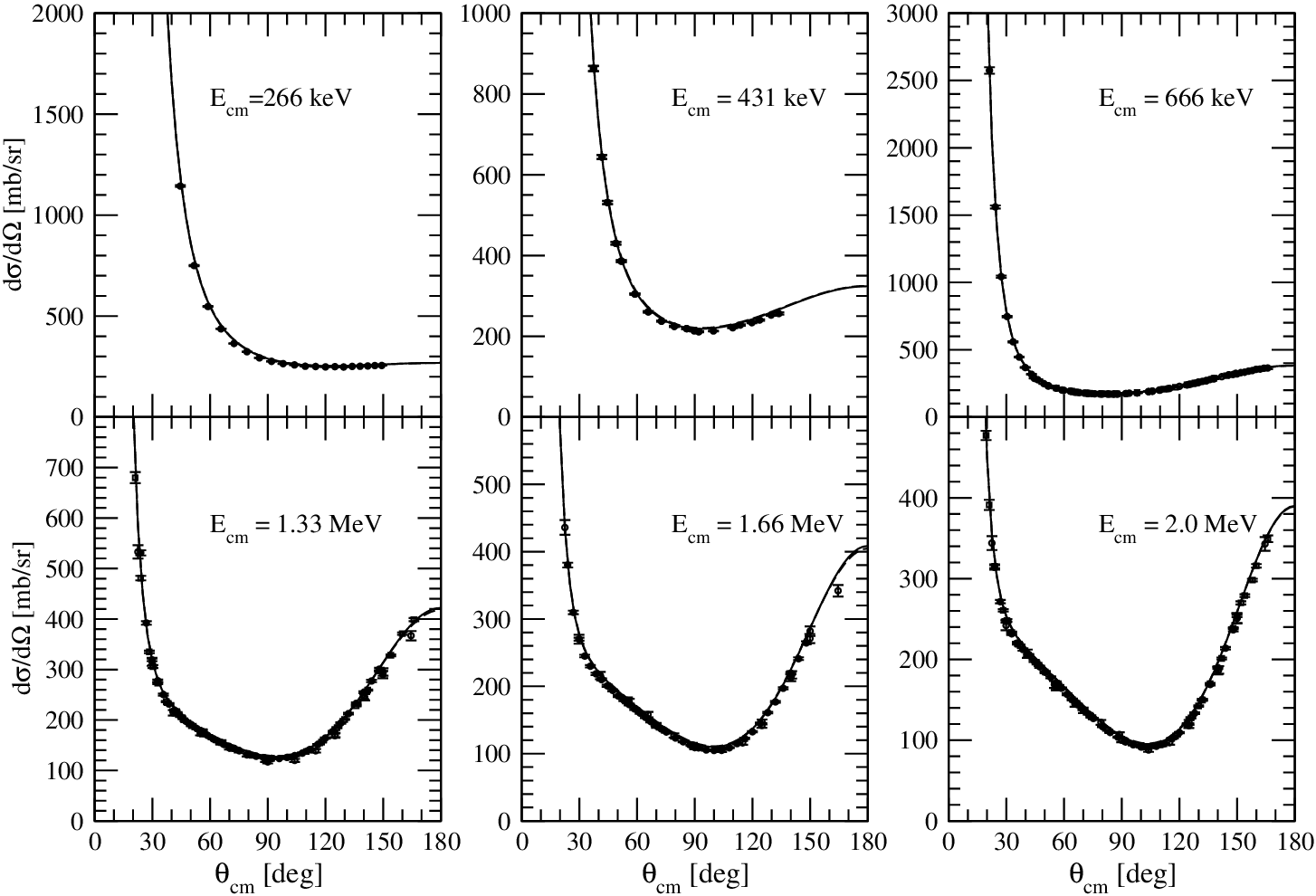}
\end{center}
\caption{\label{fig:xss2b} 
$p-d$ differential cross section for $E_{cm}=0.266, 0.431, 0.666, 
1.33, 1.66$ and 2.0 MeV calculated with the 
AV18 (dashed lines) and the N3LO-Idaho (solid lines) two-nucleon potential 
models. 
Data are from 
Ref.~\protect\cite{Hut83} at $E_{cm}=0.266$ MeV, from 
Ref.~\protect\cite{Bru01} at $E_{cm}=0.431$ MeV, from 
Refs.~\protect\cite{Hut83} (solid circles), \protect\cite{Woo01-02} 
(empty circles), and~\protect\cite{Koc69} (empty squares) 
at $E_{cm}=0.666$ MeV, from 
Refs.~\protect\cite{Koc69} (empty squares -- $E_p=1.993$ MeV), 
\protect\cite{Shi95} (solid circles), 
and~\protect\cite{She47} (empty circles -- $E_p=2.08$ MeV) 
at $E_{cm}=1.33$ MeV, from 
Refs.~\protect\cite{Shi95} (solid circles) and~\protect\cite{She47} 
(empty circles -- $E_p=2.53$ MeV) at $E_{cm}=1.66$ MeV, from 
Refs.~\protect\cite{Shi95} (solid circles), \protect\cite{Koc69} 
(empty squares -- $E_p=2.995$ MeV),
and~\protect\cite{She47} (empty circles) at $E_{cm}=2.0$ MeV.}
\end{figure}
\begin{figure}[tbp]
\begin{center}
\includegraphics[width=\textwidth,height=0.5\textheight]{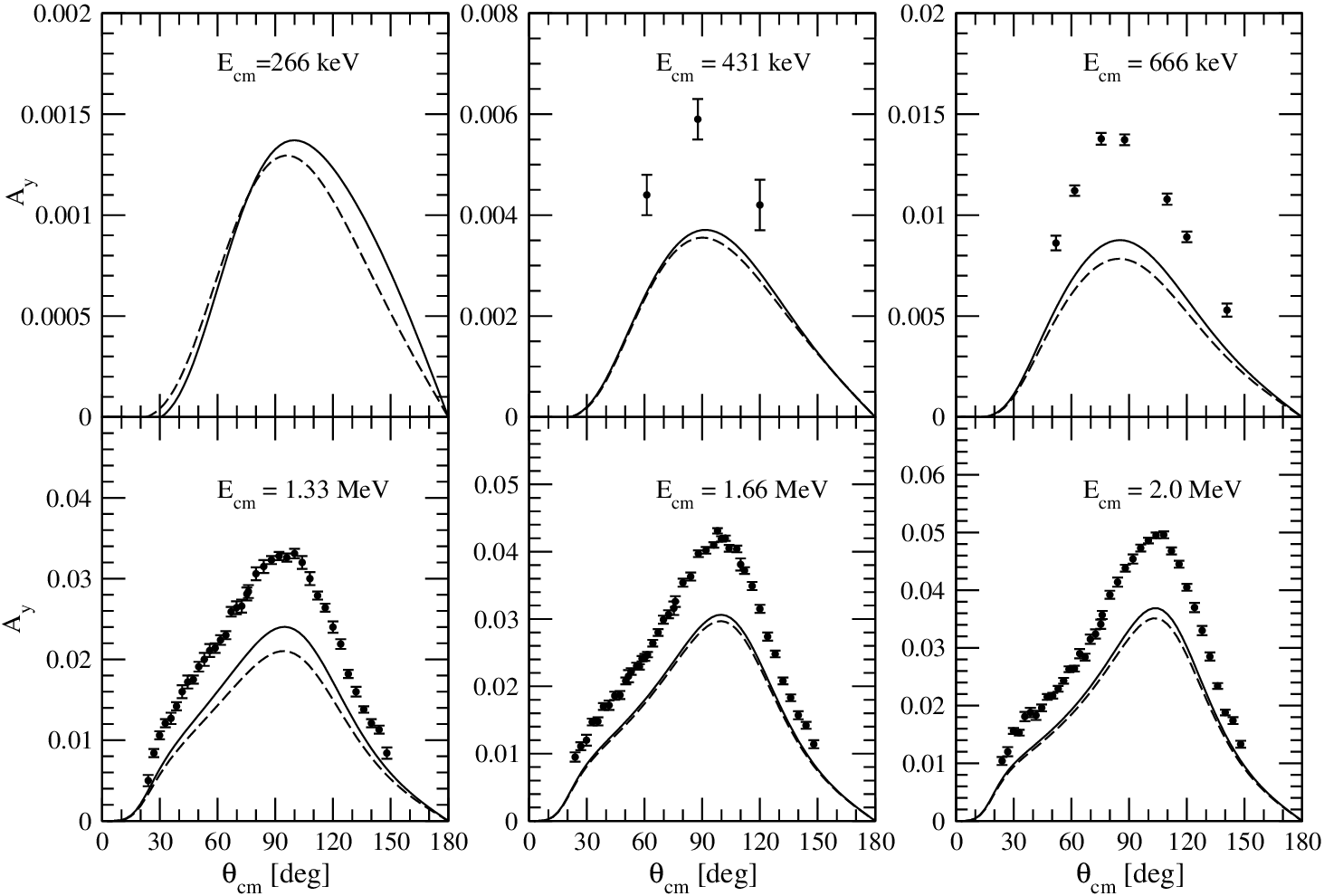}
\end{center}
\caption{\label{fig:aay2b} Same as 
Fig.~\protect\ref{fig:xss2b}, but for the proton vector 
analyzing power $A_y$. 
Data are from 
Ref.~\protect\cite{Bru01} at $E_{cm}=0.431$ MeV, from 
Ref.~\protect\cite{Woo01-02} at $E_{cm}=0.666$ MeV, from 
Ref.~\protect\cite{Shi95} at $E_{cm}=1.33, 1.66$ and 2.0 MeV.}
\end{figure}
\begin{figure}[tbp]
\begin{center}
\includegraphics[width=\textwidth,height=0.5\textheight]{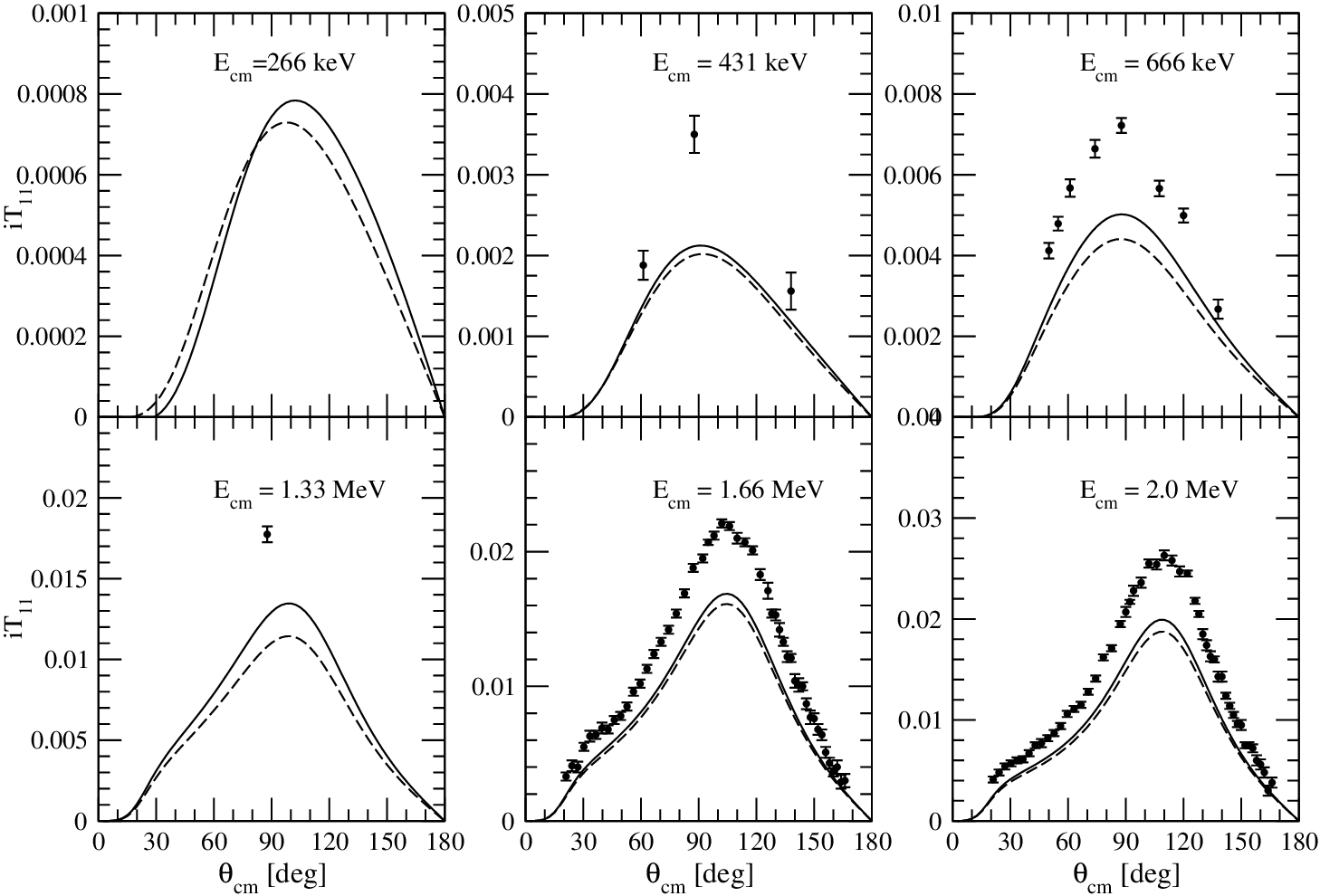}
\end{center}
\caption{\label{fig:t112b} Same as 
Fig.~\protect\ref{fig:xss2b}, but for the deuteron vector 
analyzing power $iT_{11}$. 
Data are from 
Ref.~\protect\cite{Bru01} at $E_{cm}=0.431$ and 1.33 MeV, from 
Ref.~\protect\cite{Woo01-02} at $E_{cm}=0.666$ MeV, from 
Ref.~\protect\cite{Shi95} at $E_{cm}=1.66$ and 2.0 MeV.}
\end{figure}
\begin{figure}[tbp]
\begin{center}
\includegraphics[width=\textwidth,height=0.5\textheight]{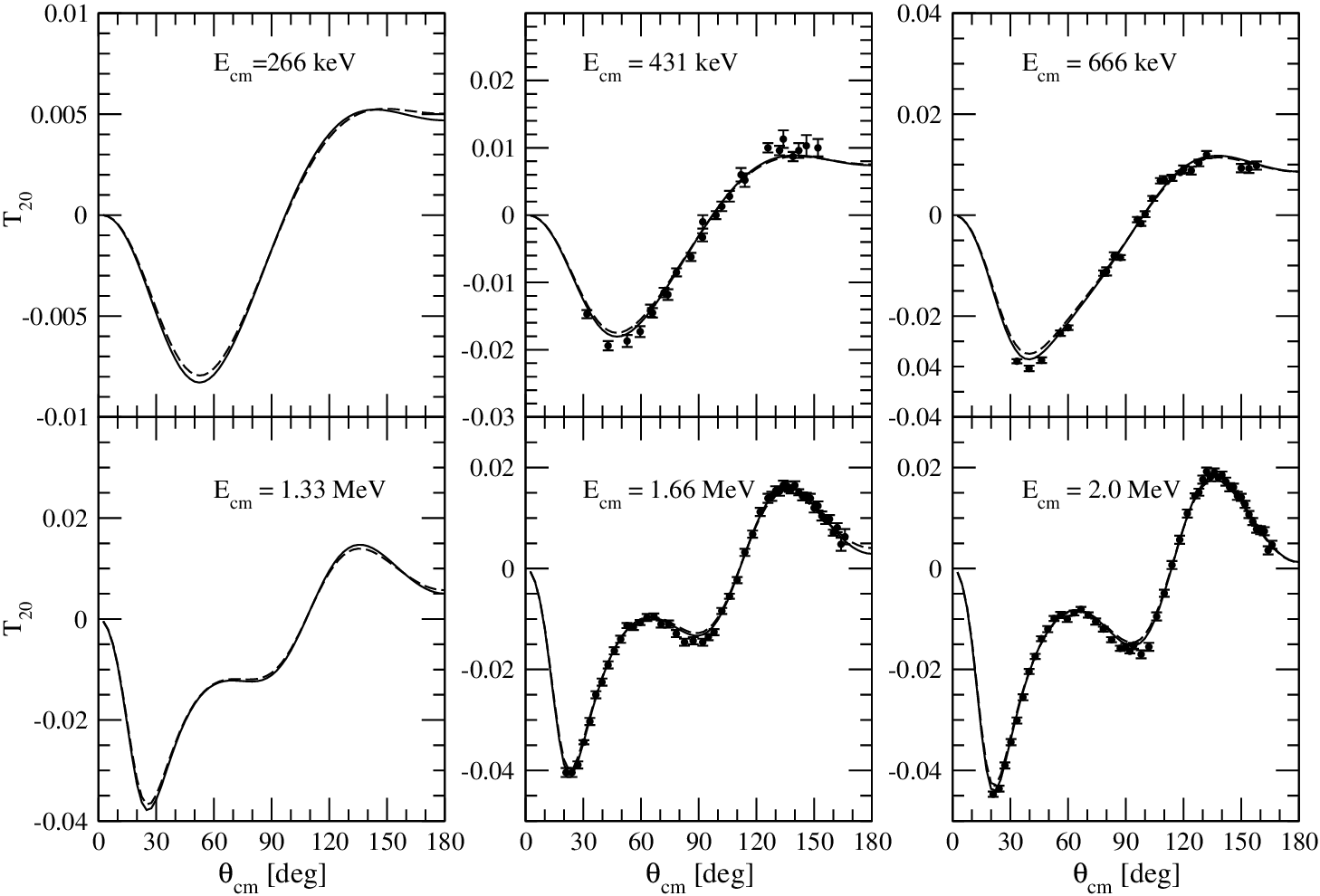}
\end{center}
\caption{\label{fig:t202b} Same as 
Fig.~\protect\ref{fig:xss2b}, but for the deuteron tensor
analyzing power $T_{20}$. 
Data are from 
Ref.~\protect\cite{Bru01} at $E_{cm}=0.431$ MeV, from 
Ref.~\protect\cite{Woo01-02} at $E_{cm}=0.666$ MeV, from 
Refs.~\protect\cite{Shi95} at $E_{cm}=1.66$ and 2.0 MeV.}
\end{figure}
\begin{figure}[tbp]
\begin{center}
\includegraphics[width=\textwidth,height=0.5\textheight]{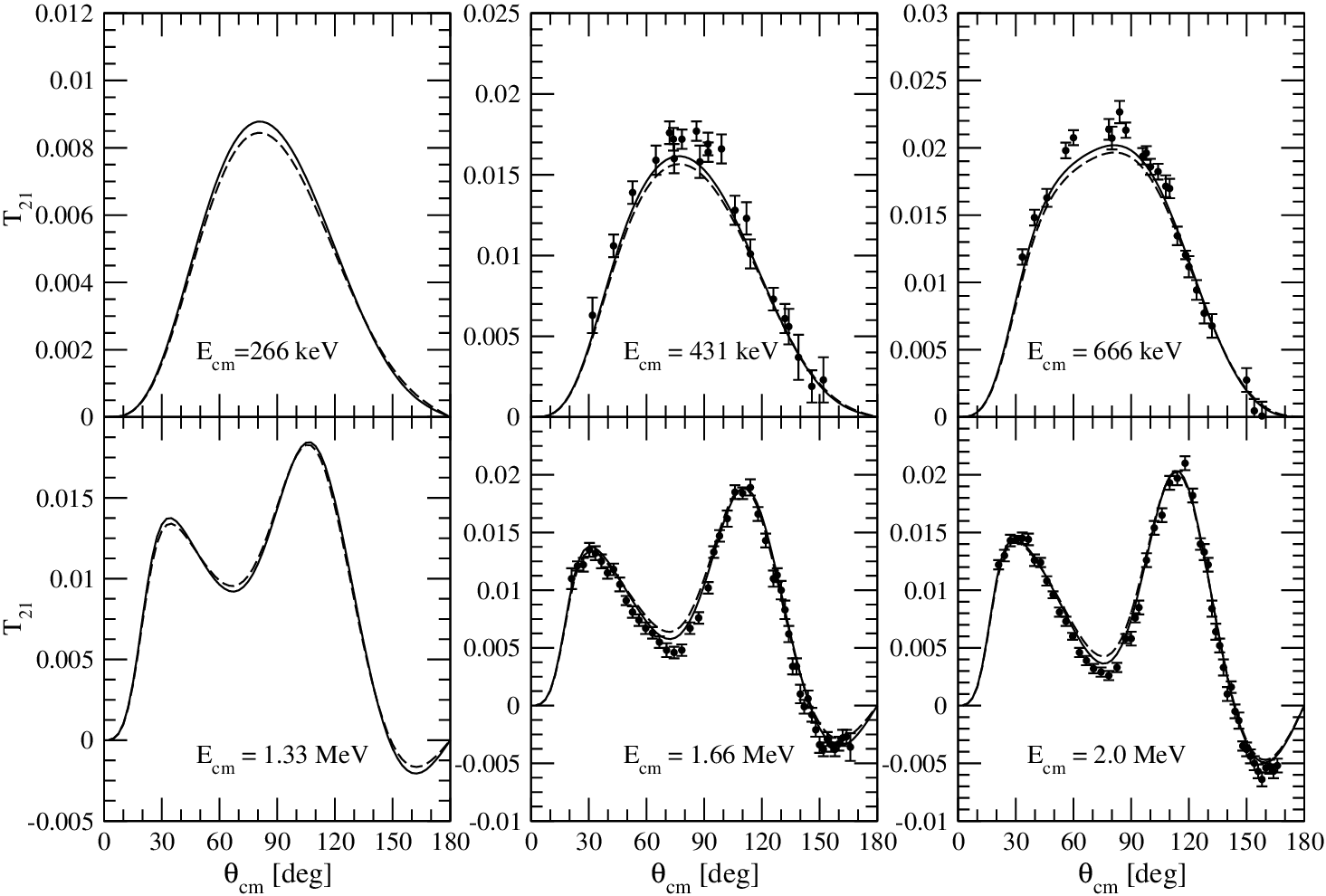}
\end{center}
\caption{\label{fig:t212b} Same as 
Fig.~\protect\ref{fig:xss2b}, but for the deuteron tensor
analyzing power $T_{21}$. 
Data are from the same references as in Fig.~\protect\ref{fig:t202b}.}
\end{figure}
\begin{figure}[tbp]
\begin{center}
\includegraphics[width=\textwidth,height=0.5\textheight]{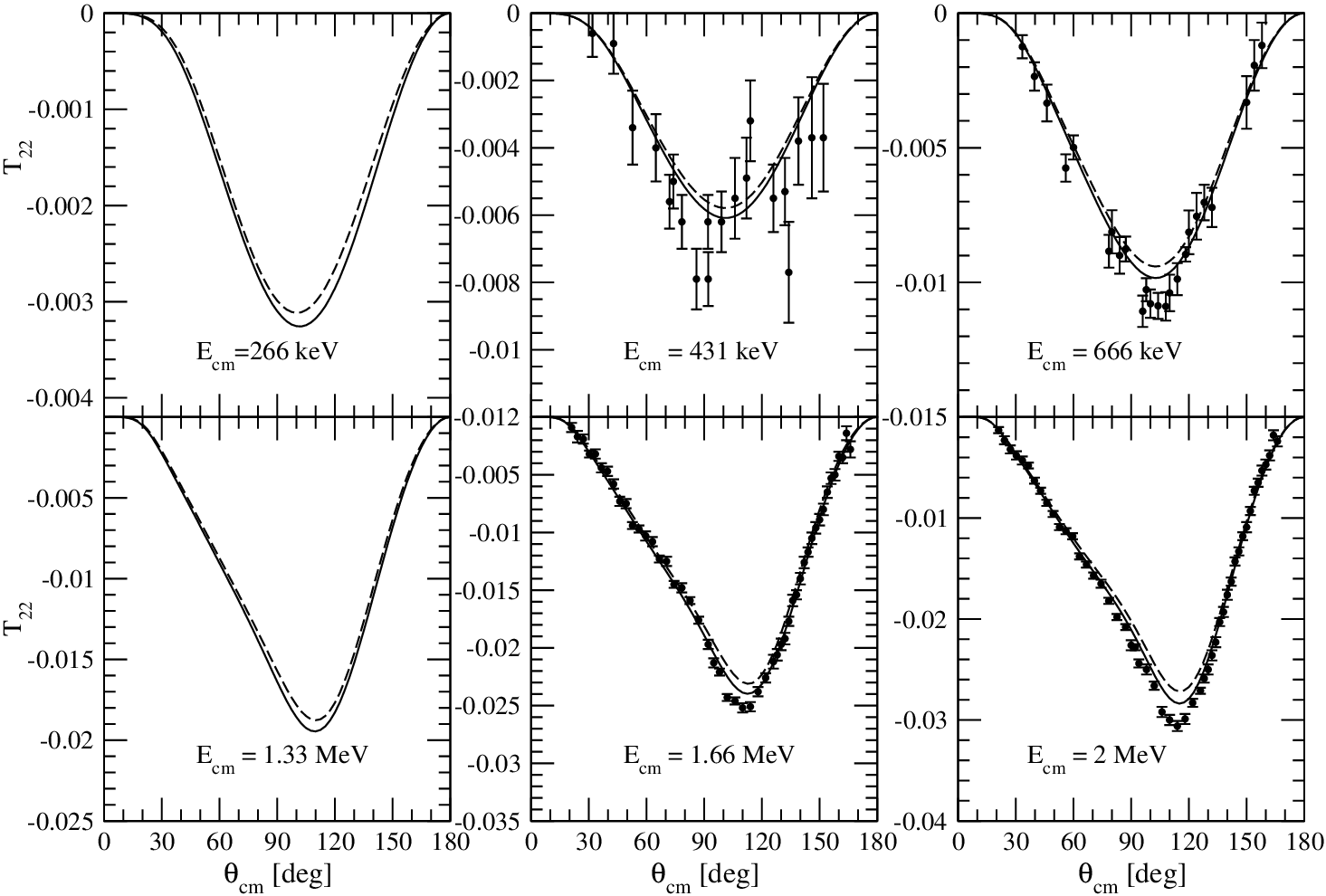}
\end{center}
\caption{\label{fig:t222b} Same as 
Fig.~\protect\ref{fig:xss2b}, but for the deuteron tensor
analyzing power $T_{22}$. 
Data are from the same references as in Fig.~\protect\ref{fig:t202b}.}
\end{figure}
The $p-d$ elastic scattering observables have been studied 
at different values of center-of-mass energy $E_{cm}$. 
Since we have considered several 
interaction models, we first focus our attention on the 
two-nucleon only models, i.e. the AV18 and the N3LO-Idaho. 
The differential 
cross section $d\sigma/d\Omega$, the proton vector analyzing power 
$A_y$, the deuteron vector and tensor 
analyzing powers $iT_{11}$, $T_{20}$, $T_{21}$ and $T_{22}$,  as 
function of the center-of-mass angle $\theta_{cm}$, are given in 
Figs.~\ref{fig:xss2b}, \ref{fig:aay2b}, \ref{fig:t112b}, \ref{fig:t202b}, 
\ref{fig:t212b} and~\ref{fig:t222b}, respectively.
The data are taken from 
Refs.~\cite{Hut83,Bru01,Woo01-02,Koc69,Shi95,She47},
as indicated in detail in the figure captions.
By inspection of the figures, we can observe that: 
(i) theory and experiment are in disagreement
for the $A_y$ and $iT_{11}$ observables 
(the well-known ``$A_y$-puzzle''~\cite{Wit94,Kie95}); 
(ii) no differences between the AV18 and 
the N3LO-Idaho curves can be seen for the 
differential cross sections; (iii) 
the N3LO-Idaho curves are systematically closer to the data 
than the AV18 ones for the polarization observables, 
especially for $A_y$ and $iT_{11}$. The reason of this behaviour is 
well known~\cite{Kie04} and is related to the MM interaction.
In fact, the AV18 potential model has been constructed keeping 
the electromagnetic interaction separated from the nuclear one. The 
electromagnetic interaction includes the MM one, as well as 
higher-order corrections to the $pp$ Coulomb potential as two-photon exchange, 
Darwin-Foldy and vacuum polarization terms. The MM interaction effects 
are known to be sizable in $N-d$ elastic scattering~\cite{Kie04}.
On the contrary, the N3LO-Idaho potential 
model keeps as electromagnetic interaction only the point Coulomb 
potential and MM effects are indirectly included in the 
nuclear part of the interaction by the fitting procedure. 
From this observation, 
we can guess that the results obtained with the two-nucleon potentials
AV18 and N3LO-Idaho should be comparable when the AV18 calculation includes 
also the MM effects. To verify this hypothesis, we have calculated the $p-d$
elastic scattering observables at two values of $E_{cm}$, 1.33 and 
2.0 MeV, using the AV18, AV18+MM, and N3LO-Idaho potential models. 
The results are given in Figs.~\ref{fig:mm1.33} and~\ref{fig:mm2}, 
respectively. From inspection of the figures, we can notice that
the AV18+MM results for the $A_y$ and $iT_{11}$ vector polarization 
observables are larger than the AV18 alone ones in the 
maximum region, and that the AV18+MM
and N3LO-Idaho curves are quite close to each other 
for all the observables considered. Although this analysis should be 
performed systematically at any value of $E_{cm}$ and for any observable,
given the conclusions of Ref.~\cite{Kie04}, 
it can be expected that a similar behaviour still holds. 
Therefore, we can conclude that the non-local N3LO-Idaho and the 
local AV18 two-nucleon interactions give similar results once the MM 
effects are included in the AV18 calculation. For this reason, we have 
chosen to use the N3LO-Idaho two-nucleon interaction model 
in the continuation of our study.
\begin{figure}[tbp]
\begin{center}
\includegraphics[width=0.85\textwidth,height=0.75\textheight]{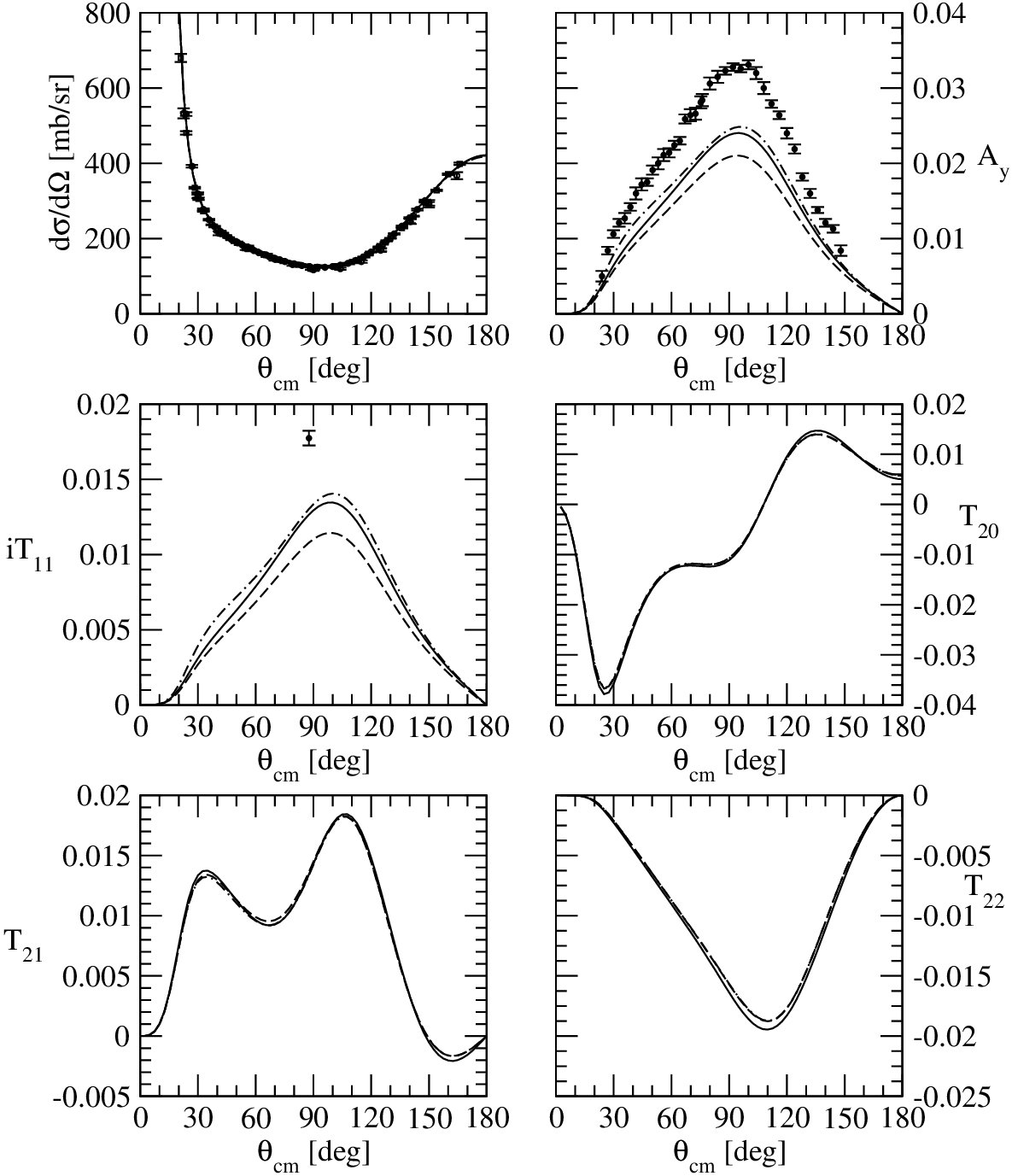}
\end{center}
\caption{\label{fig:mm1.33} 
Theoretical results for $p-d$ differential 
cross section $d\sigma/d\Omega$, and polarization 
observables $A_y$, $iT_{11}$, $T_{20}$, $T_{21}$ and $T_{22}$,  at 
$E_{cm}=1.33$ MeV are compared to the 
experimental data.
The calculation are done using the AV18 (dashed lines), 
the AV18+MM (dotted-dashed lines), and the N3LO-Idaho (solid lines)
interactions. The data are from 
Refs.~\protect\cite{Koc69} (empty squares -- $E_p=1.993$ MeV), 
\protect\cite{Shi95} (solid circles), 
and~\protect\cite{She47} (empty circles -- $E_p=2.08$ MeV) for the 
differential cross section, and from Refs.~\protect\cite{Shi95} 
and~\cite{Bru01} 
for the $A_y$ and $iT_{11}$ polarization observables, respectively. 
The incident proton (deuteron) 
is $E_p=2.0$ MeV ($E_d=4.0$ MeV).}
\end{figure}
\begin{figure}[tbp]
\begin{center}
\includegraphics[width=0.85\textwidth,height=0.75\textheight]{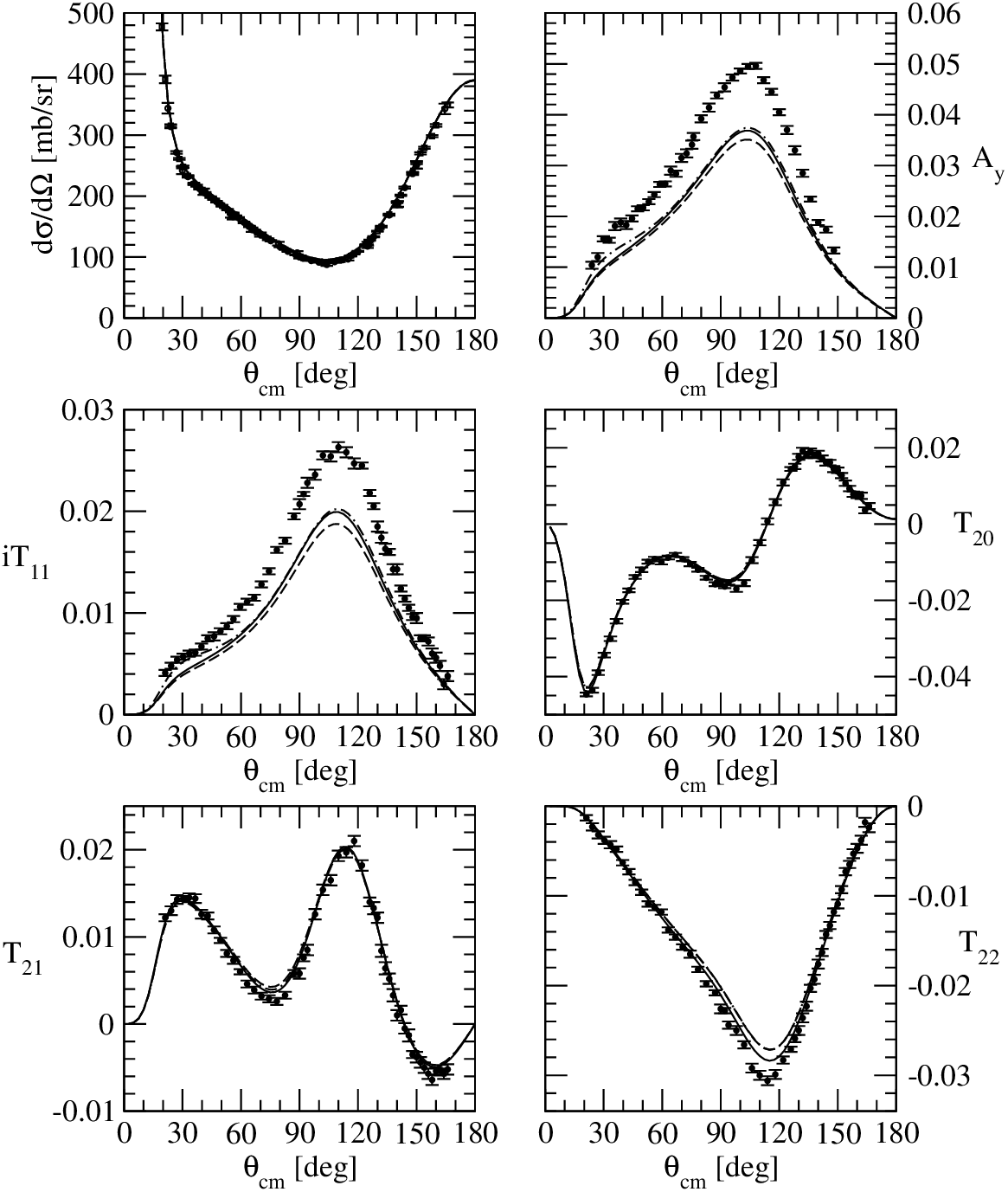}
\end{center}
\caption{\label{fig:mm2} 
Same as Fig.~\protect\ref{fig:mm1.33} but for 
$E_{cm}=2.0$ MeV.
The data are from 
Refs.~\protect\cite{Shi95} (solid circles), \protect\cite{Koc69} 
(empty squares -- $E_p=2.995$ MeV),
and~\protect\cite{She47} (empty circles) for the 
differential cross section, and from Ref.~\protect\cite{Shi95} 
for the polarization observables. The incident proton (deuteron) 
is $E_p=3.0$ MeV ($E_d=6.0$ MeV).}
\end{figure}
\begin{figure}[tbp]
\begin{center}
\includegraphics[width=\textwidth,height=0.5\textheight]{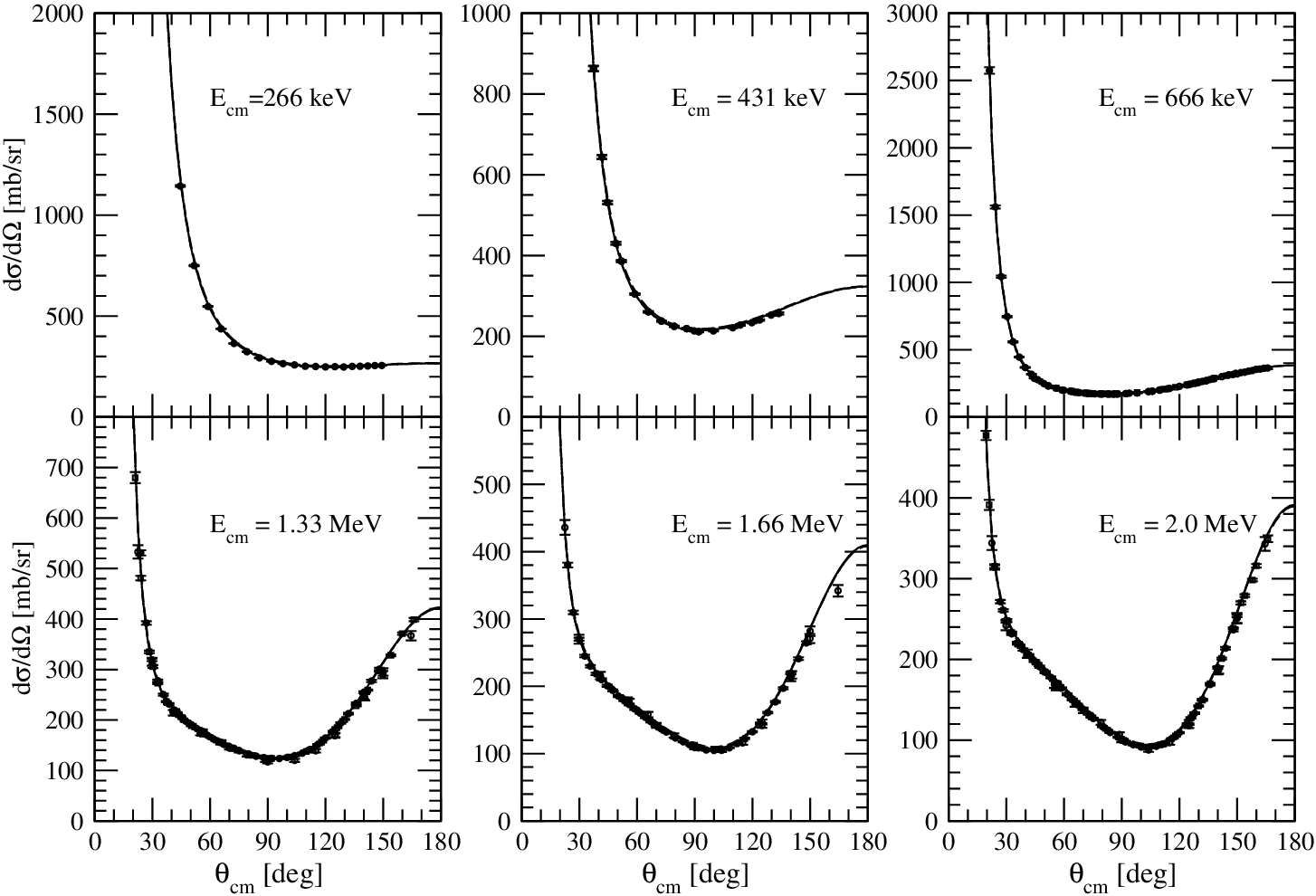}
\end{center}
\caption{\label{fig:xss} 
$p-d$ differential cross section for $E_{cm}=0.266, 0.431, 0.666, 
1.33, 1.66$ and 2.0 MeV calculated with the 
N3LO-Idaho (dashed lines), the N3LO-Idaho/UIXp (dotted-dashed lines) 
and the N3LO-Idaho/N2LO (solid lines) two- and three-nucleon interaction 
models. 
Data are from the same references as in Fig.~\protect\ref{fig:xss2b}.  
}
\end{figure}
\begin{figure}[tbp]
\begin{center}
\includegraphics[width=\textwidth,height=0.5\textheight]{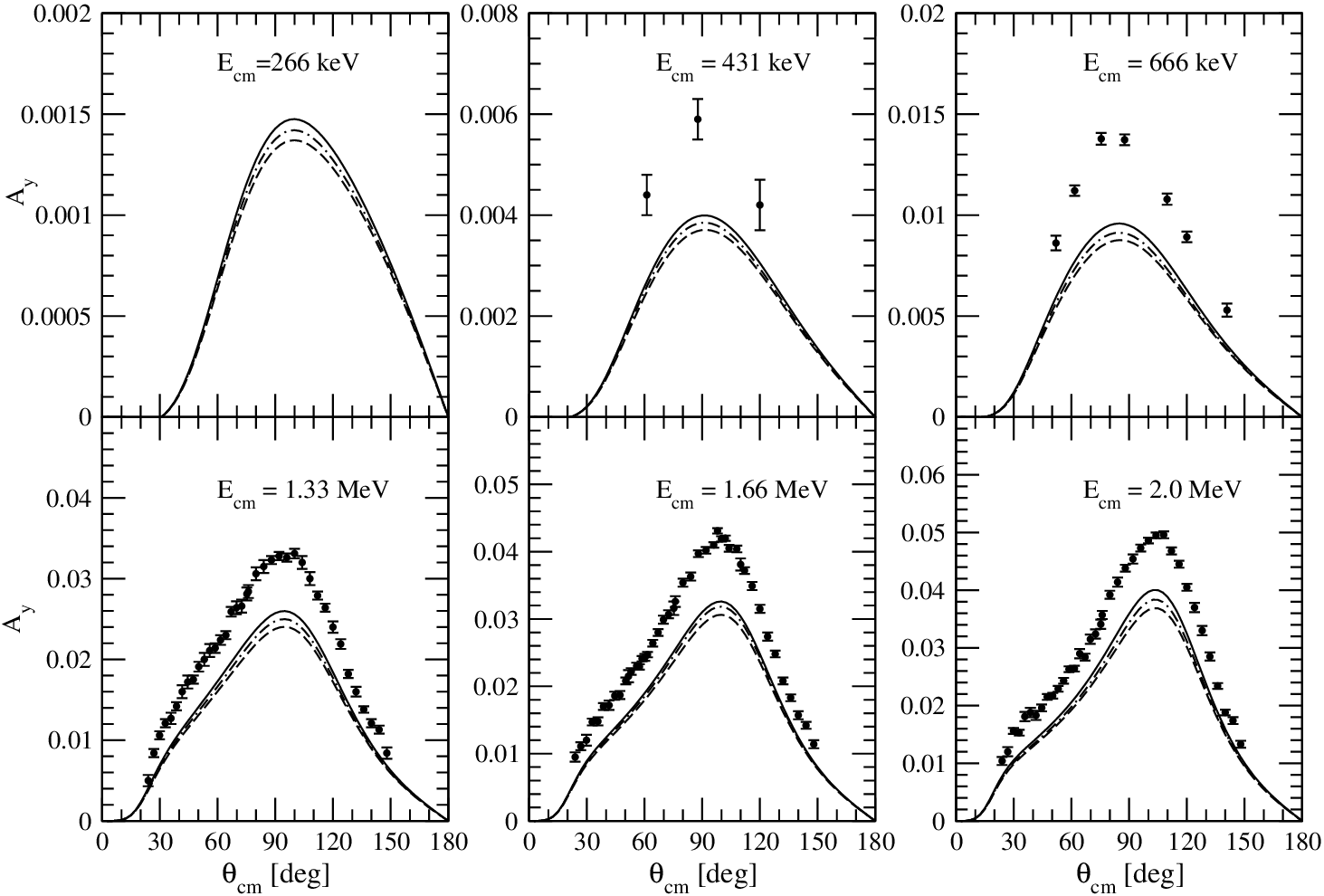}
\end{center}
\caption{\label{fig:aay} Same as 
Fig.~\protect\ref{fig:xss}, but for the proton vector 
analyzing power $A_y$. 
Data are from the same references as in Fig.~\protect\ref{fig:aay2b}.  
}
\end{figure}
\begin{figure}[tbp]
\begin{center}
\includegraphics[width=\textwidth,height=0.5\textheight]{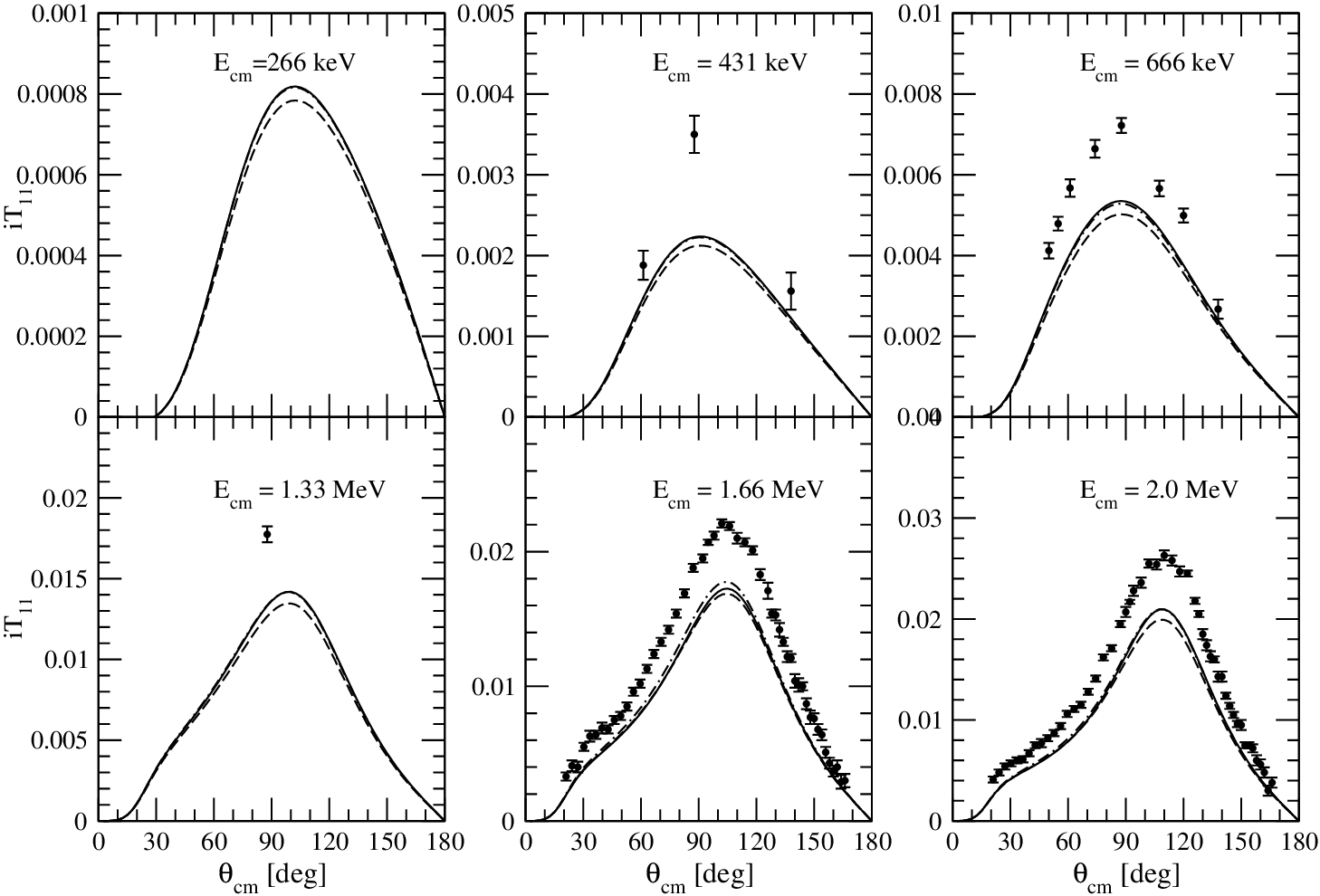}
\end{center}
\caption{\label{fig:t11} Same as 
Fig.~\protect\ref{fig:xss}, but for the deuteron vector 
analyzing power $iT_{11}$. 
Data are from the same references as in Fig.~\protect\ref{fig:t112b}.  
}
\end{figure}
\begin{figure}[tbp]
\begin{center}
\includegraphics[width=\textwidth,height=0.5\textheight]{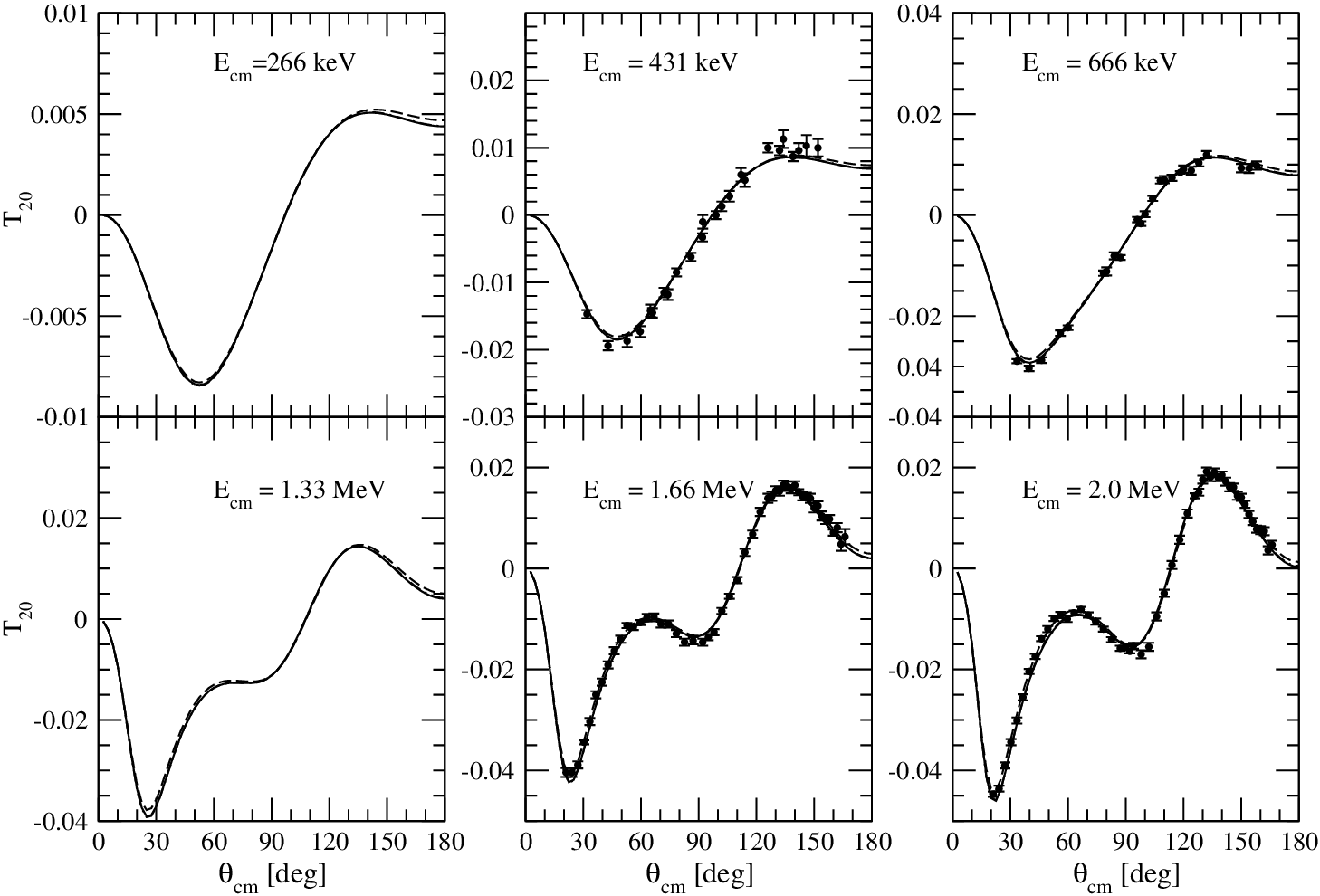}
\end{center}
\caption{\label{fig:t20} Same as 
Fig.~\protect\ref{fig:xss}, but for the deuteron tensor
analyzing power $T_{20}$. 
Data are from the same references as in Fig.~\protect\ref{fig:t202b}.  
}
\end{figure}
\begin{figure}[tbp]
\begin{center}
\includegraphics[width=\textwidth,height=0.5\textheight]{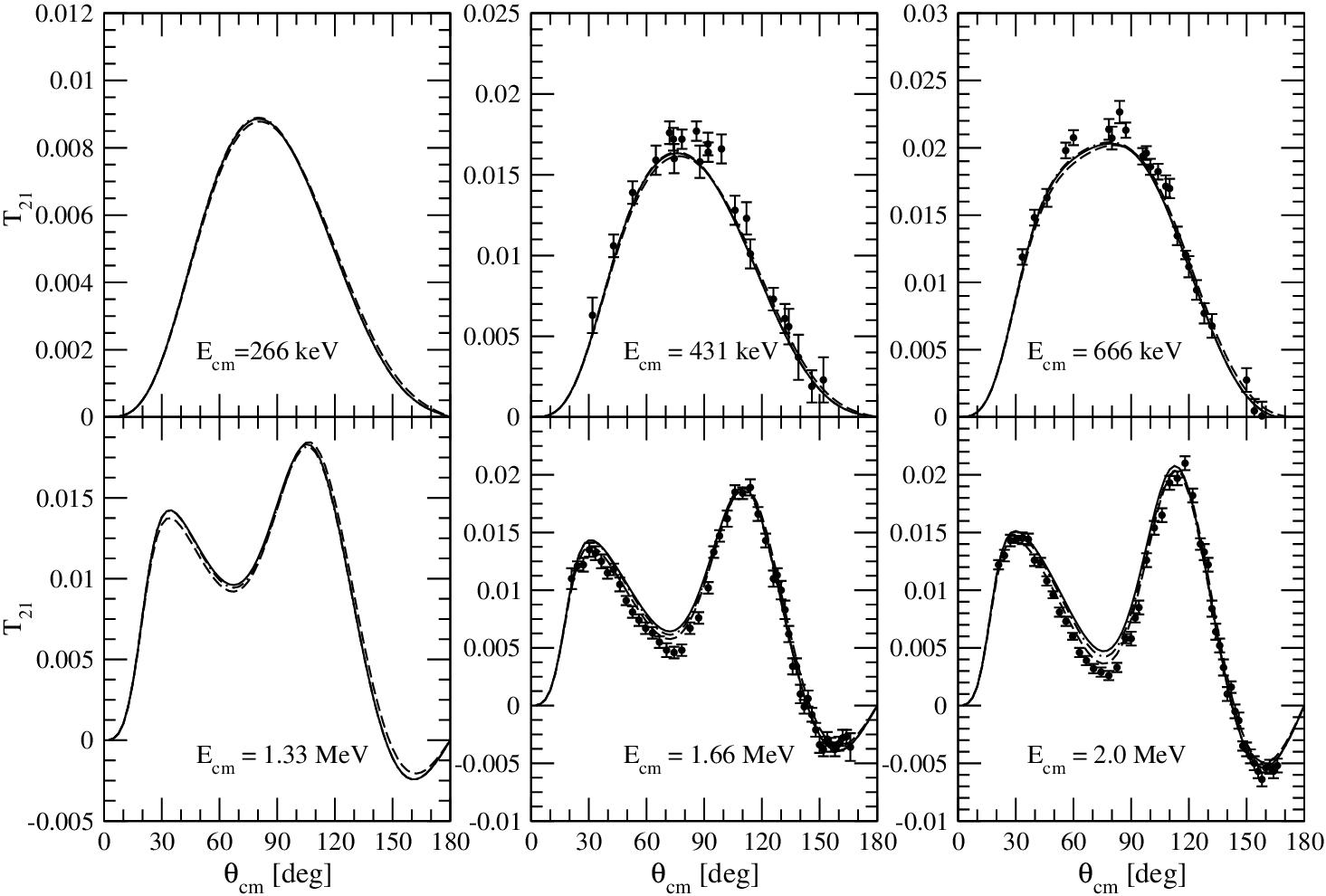}
\end{center}
\caption{\label{fig:t21} Same as 
Fig.~\protect\ref{fig:xss}, but for the deuteron tensor
analyzing power $T_{21}$. 
Data are from the same references as in Fig.~\protect\ref{fig:t212b}.}
\end{figure}
\begin{figure}[tbp]
\begin{center}
\includegraphics[width=\textwidth,height=0.5\textheight]{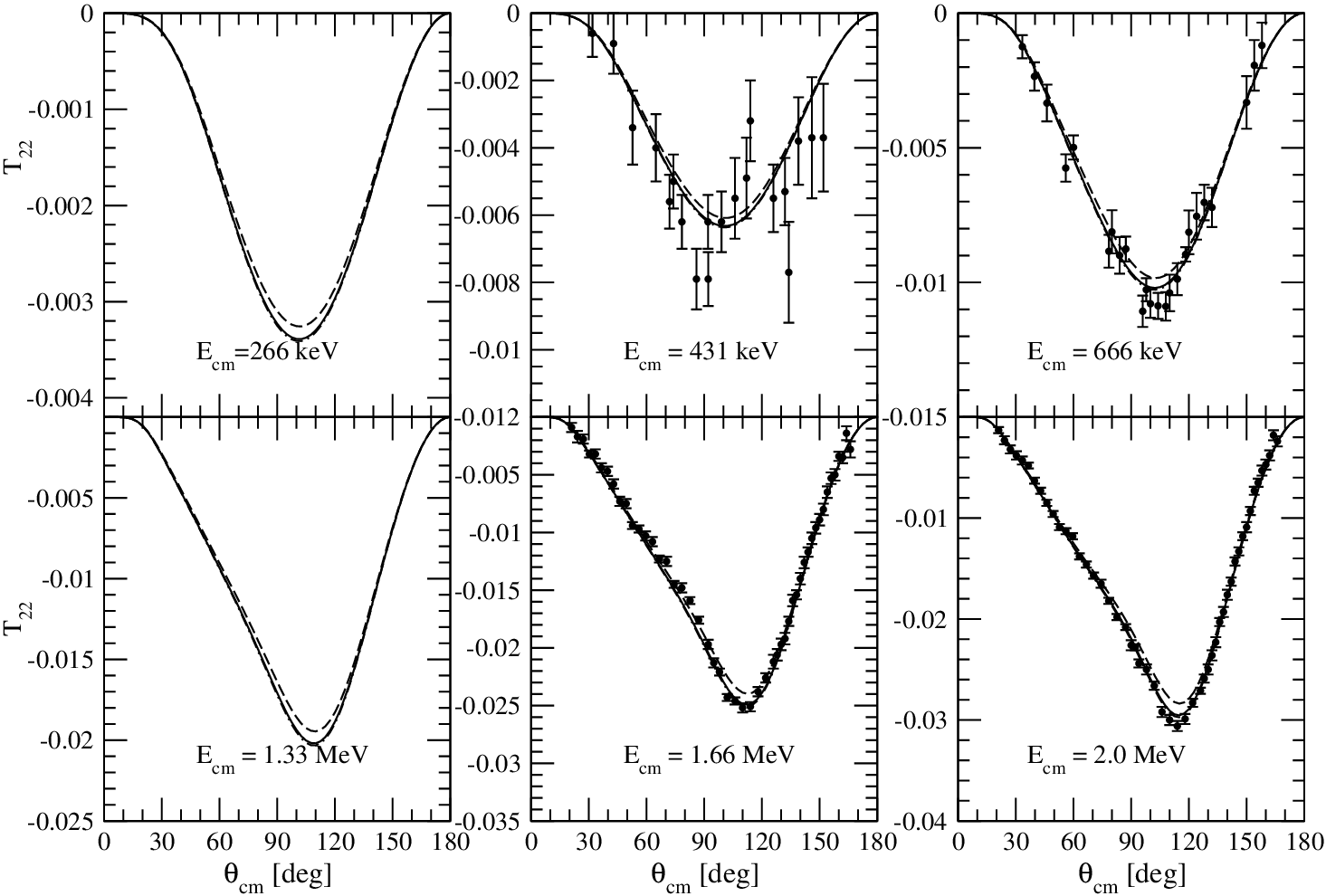}
\end{center}
\caption{\label{fig:t22} Same as 
Fig.~\protect\ref{fig:xss}, but for the deuteron tensor
analyzing power $T_{22}$. 
Data are from the same references as in Fig.~\protect\ref{fig:t222b}.}
\end{figure}

In order to have a meaningful comparison with the data, 
the TNI cannot be neglected in the calculation.
Therefore, we present in Figs.~\ref{fig:xss}, \ref{fig:aay}, \ref{fig:t11}, 
\ref{fig:t20}, \ref{fig:t21} and~\ref{fig:t22} the results for the 
different observables, obtained with the N3LO-Idaho two-nucleon,  
and the N3LO-Idaho/UIXp 
and N3LO-Idaho/N2LO two- and three-nucleon interaction models. From inspection 
of the figures, we can observe that 
the TNI effects are sizable, especially for the polarization observables, 
and the N3LO-Idaho/N2LO potential 
model gives a slightly better description of the data than 
the N3LO-Idaho/UIXp one. In particular, it is interesting to notice 
that the $A_y$ and $iT_{11}$ observables are better described 
at every value of $E_{cm}$, except for 
$iT_{11}$ at $E_{cm}=1.66$ MeV, although even in this case all the curves 
are very close to each other. 

For a better comparison between the different 
potential models and the data, a $\chi^2$ analysis has been carried 
only for those
observables, except the differential cross section, for which 
the number of data $N$ is $N\geq 7$. In particular, 
following Ref.~\cite{Kie01a}, 
\begin{equation}
\chi^2/{\rm datum}=\frac{1}{N}\sum_i\frac{(f_i^{exp}-f_i^{th})^2}              
{(\Delta f_i)^2} \ ,
\label{eq:chi2}
\end{equation}
where $f_i^{exp}$ is the $i$th datum at center-of-mass angle $\theta_i$, 
$\Delta f_i$ is its experimental error, and $f_i^{th}$ is the theoretical
value at the same angle. 
The results are given in Table~\ref{tab:chi2} for 
$E_{cm}=0.666, 1.33, 1.66$ and 2.0 MeV. 
The N3LO-Idaho, N3LO-Idaho/UIXp and N3LO-Idaho/N2LO interaction 
models have been considered. 
From inspection of the table we can notice that all the values for 
$\chi^2$/datum are comparable, although the ones obtained with 
the N3LO-Idaho two-nucleon interaction are usually higher than the ones 
obtained with two- and three-nucleon interactions, except for 
the tensor analyzing power $T_{20}$ and $T_{21}$. 
This is a well-known and still unclear issue, i.e.\  
$T_{20}$ and $T_{21}$ are 
better described, as the energy increases, 
by two-nucleon only interaction models, even at 30.0 MeV~\cite{Kie01b}.
Among the two- plus three-nucleon 
interaction models, the N3LO-Idaho/N2LO performs slightly better than 
the N3LO-Idaho/UIXp. 
\begin{table}
\caption{\label{tab:chi2}
$\chi^2$/datum of the $p-d$ elastic scattering observables
at $E_{cm}=0.666, 1.33, 1.66$ and 2.0 MeV, 
calculated with the N3LO-Idaho 
two-nucleon only, and the N3LO-Idaho/UIXp and 
N3LO-Idaho/N2LO two- plus three-nucleon Hamiltonian models. 
The different number $N$ of experimental data is also indicated.
The data are from Ref.~\protect\cite{Woo01-02} at $E_{cm}=0.666$ MeV, 
and from Ref.~\cite{Shi95} at $E_{cm}=1.33, 1.66$ and 2.0 MeV.}
\begin{center}
\begin{tabular}{c|ccccc|c|ccccc|ccccc}
\hline
& \multicolumn{5}{c|}{0.666 MeV} & 1.33 MeV 
& \multicolumn{5}{c|}{1.66 MeV} & \multicolumn{5}{c}{2.0 MeV} \\
& $A_y$ & $iT_{11}$ & $T_{20}$ & $T_{21}$ & $T_{22}$ 
& $A_y$ 
& $A_y$ & $iT_{11}$ & $T_{20}$ & $T_{21}$ & $T_{22}$ 
& $A_y$ & $iT_{11}$ & $T_{20}$ & $T_{21}$ & $T_{22}$ \\
\hline
$N$ & 7 & 8 & 24 & 24 & 24 
& 38 
& 44 & 50 & 50 & 50 & 50 
& 38 & 51 & 51 & 51 & 51   \\
\hline
N3LO-Idaho & 197.7 & 68.7 & 4.0 & 2.6 &  1.5 
& 108.4 
& 227.9 & 92.6 & 1.0 & 2.2 &  2.7 
& 186.0 & 108.3 & 1.9 & 2.8 &  4.4 \\
N3LO-Idaho/UIXp & 171.2 & 53.1 & 2.6 & 2.2 & 0.9 
& 89.2
& 185.9 & 67.0  & 2.0 & 3.2 & 3.2 
& 152.5 & 81.8 & 3.0 & 5.5 & 1.6 \\
N3LO-Idaho/N2LO & 139.9 & 49.5 & 2.7 & 2.5 & 0.9 
& 70.0
& 159.4 & 84.3 & 2.1 & 4.0 & 2.8 
& 114.0 & 85.8 & 3.6 & 8.3 & 1.6 \\
\hline
\end{tabular}
\end{center}
\end{table}

The $p-d$ elastic scattering observables at $E_{cm}=0.666$ and 2.0 MeV 
have been calculated also using the two-nucleon only potential 
model $V_{low\!-\!k}$, obtained
from the AV18 with a cutoff parameter $\Lambda$ equal to 2.2 fm$^{-1}$, 
as already used 
for the calculation of the scattering lengths. The results are 
given in Figs.~\ref{fig:vlowk666keV} and~\ref{fig:vlowk2MeV}, 
respectively. Together with the $V_{low\!-\!k}$ results, 
we have shown also the bare AV18 and the 
AV18/UIX ones. From inspection of the figures, we can observe that 
the $V_{low\!-\!k}$ results are very similar to the AV18/UIX ones. 
This can be understood noticing that the considered observables are 
sensitive to $P$- and $D$-wave scattering. 
The $P$-wave phase shifts and mixing angles are influenced 
by the UIX TNI attraction term, which is reproduced, 
within the $V_{low\!-\!k}$ 
approach, by fitting the cutoff parameter $\Lambda$. 
In fact, 
the $J^\pi=1/2^-$ phase shifts and mixing angle 
$(\delta_{1\frac{1}{2}\frac{1}{2}}, \delta_{1\frac{3}{2}\frac{1}{2}}, 
\epsilon_{{\frac{1}{2}}^-})$ obtained 
at $E_{cm}=2.0$ MeV
with the AV18, AV18/UIX and $V_{low\!-\!k}$ potential models 
are $(-7.358,22.11,5.718)$, $(-7.366,22.32,5.835)$, 
and $(-7.343,22.26,5.811)$, respectively. 
From this 
first analysis of $V_{low\!-\!k}$ results for $N-d$ scattering
at low energies, 
we can conclude that the $V_{low\!-\!k}$ and AV18/UIX results 
are close to each other for observables sensitive to $P$-
and $D$-wave
scattering, like vector and tensor 
analyzing powers. Further work on these aspects 
is currently underway.
\begin{figure}[tbp]
\begin{center}
\includegraphics[width=0.85\textwidth,height=0.75\textheight]{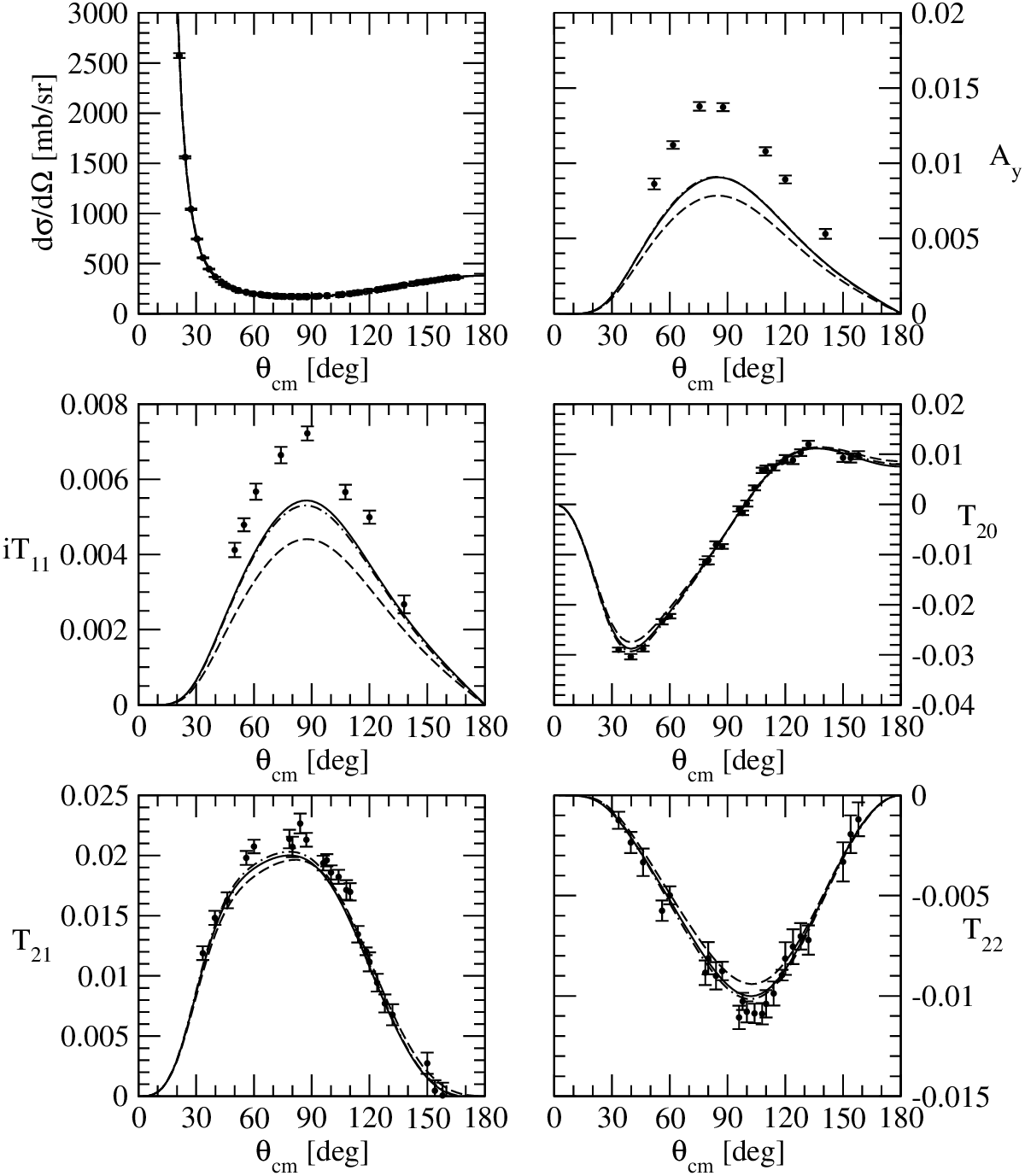}
\end{center}
\caption{\label{fig:vlowk666keV} 
The theoretical values of the $p-d$ differential 
cross section $d\sigma/d\Omega$, 
$A_y$, $iT_{11}$, $T_{20}$, $T_{21}$ and $T_{22}$,  at 
$E_{cm}=0.666$ MeV are compared to the 
experimental data.
The calculation are done with the AV18 (dashed lines), the AV18/UIX 
(dotted-dashed lines) and the 
the $V_{low\!-\!k}$ two-nucleon interaction, 
obtained from the AV18 
with a cutoff parameter $\Lambda=2.2$ fm$^{-1}$ (solid lines).
The data are from Refs.~\protect\cite{Hut83} (solid circles),
\protect\cite{Woo01-02} (empty circles) and~\protect\cite{Koc69} 
(empty squares) for the 
differential cross section, and from Ref.~\protect\cite{Woo01-02} 
for the polarization observables. The incident proton (deuteron) 
is $E_p=1.0$ MeV ($E_d=2.0$ MeV).}
\end{figure}
\begin{figure}[tbp]
\begin{center}
\includegraphics[width=0.85\textwidth,height=0.75\textheight]{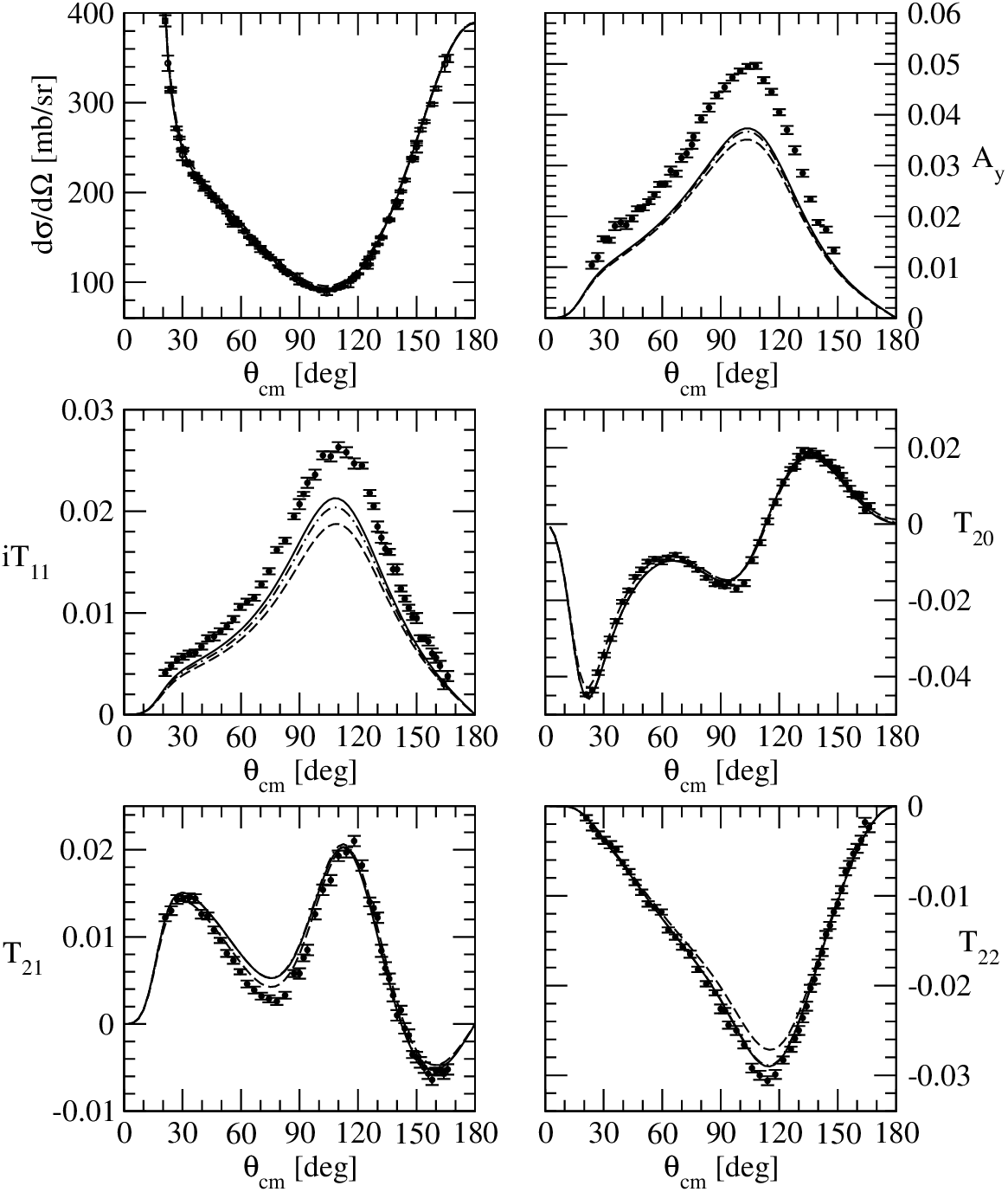}
\end{center}
\caption{\label{fig:vlowk2MeV} Same as Fig.~\protect\ref{fig:vlowk666keV}, 
but at $E_{cm}=2.0$ MeV . 
The data are from Refs.~\protect\cite{Shi95} (solid circles),
\protect\cite{Koc69} (empty circles) and~\protect\cite{She47} 
(empty squares) for the 
differential cross section, and from Ref.~\protect\cite{Shi95} 
for the polarization observables. The incident proton (deuteron) 
is $E_p=3.0$ MeV ($E_d=6.0$ MeV).
}
\end{figure}

The $n-d$ elastic scattering observables, including differential 
cross section, neutron vector analyzing power $A_y$, deuteron 
vector and tensor 
analyzing powers $iT_{11}$, $T_{20}$, $T_{21}$ and $T_{22}$,  at 
$E_{cm}=1.33$ and 2.0 MeV are 
given in Figs.~\ref{fig:ndobs1.33} and \ref{fig:ndobs2}, respectively.
The experimental data are from Refs.~\cite{Ada53,Web81,Nei03}
and Refs.~\cite{Sch83,McA94} at $E_{cm}=1.33$ MeV and 2.0 MeV, respectively.
The different curves are obtained using the N3LO-Idaho,  
N3LO-Idaho/UIXp and N3LO-Idaho/N2LO potential models.
From inspection of the figures, we can observe that all the curves 
are very close to each other, especially for the differential cross section
$d\sigma/d\Omega$ and the tensor analyzing powers 
$T_{20}$, $T_{21}$ and $T_{22}$, although some small differences are 
appreciable. Moreover, some differences are present for the 
$A_y$ and $iT_{11}$ vector polarization observables at the peak, 
even if TNI effects are small. 
Comparing the calculations with the data, we can 
observe that the calculated $d\sigma/d\Omega$ at $E_{cm}=1.33$ MeV 
is much lower than the measured one 
for large values of the center-of-mass angle $\theta_{cm}$. 
Such a discrepancy however disappears at $E_{cm}=2.0$ MeV. This difference 
has been observed before and its origin 
has still to be clarified~\cite{Kie96}. As in the $p-d$ case, 
the $n-d$ vector analyzing powers 
$A_y$ are poorly described by the theory 
in the maximum region,
but it should be 
noticed that the N3LO-Idaho/N2LO gives again a better description 
of the observables than the N3LO-Idaho/UIXp Hamiltonian model. 
\begin{figure}[tbp]
\begin{center}
\includegraphics[width=0.85\textwidth,height=0.75\textheight]{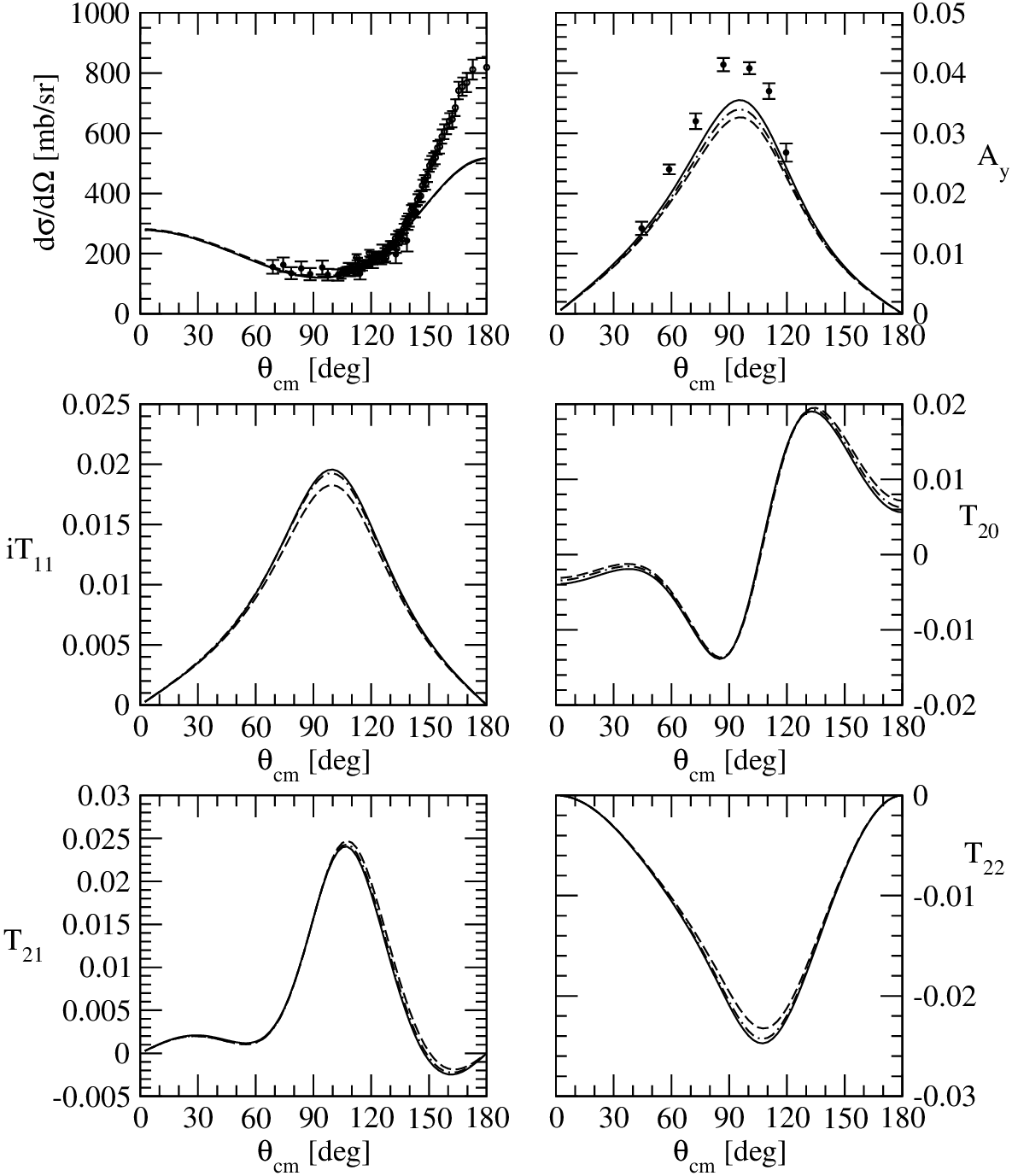}
\end{center}
\caption{\label{fig:ndobs1.33}
$n-d$ differential 
cross section $d\sigma/d\Omega$, 
$A_y$, $iT_{11}$, $T_{20}$, $T_{21}$ and $T_{22}$,  at 
$E_{cm}=1.33$ MeV are 
calculated with the N3LO-Idaho (dashed line), 
the N3LO-Idaho/UIXp (dotted-dashed line), and the N3LO-Idaho/N2LO 
(solid line) potential models.
The experimental data are of Refs.~{\protect\cite{Ada53}} (solid circles) 
and~\protect\cite{Web81} (empty squares -- $E_n=2.016$ MeV) for 
$d\sigma/d\Omega$, and Ref.~\protect\cite{Nei03} for $A_y$.
The incident neutron (deuteron) is $E_n=2.0$ MeV ($E_d=4.0$ MeV).}
\end{figure}
\begin{figure}[tbp]
\begin{center}
\includegraphics[width=0.85\textwidth,height=0.75\textheight]{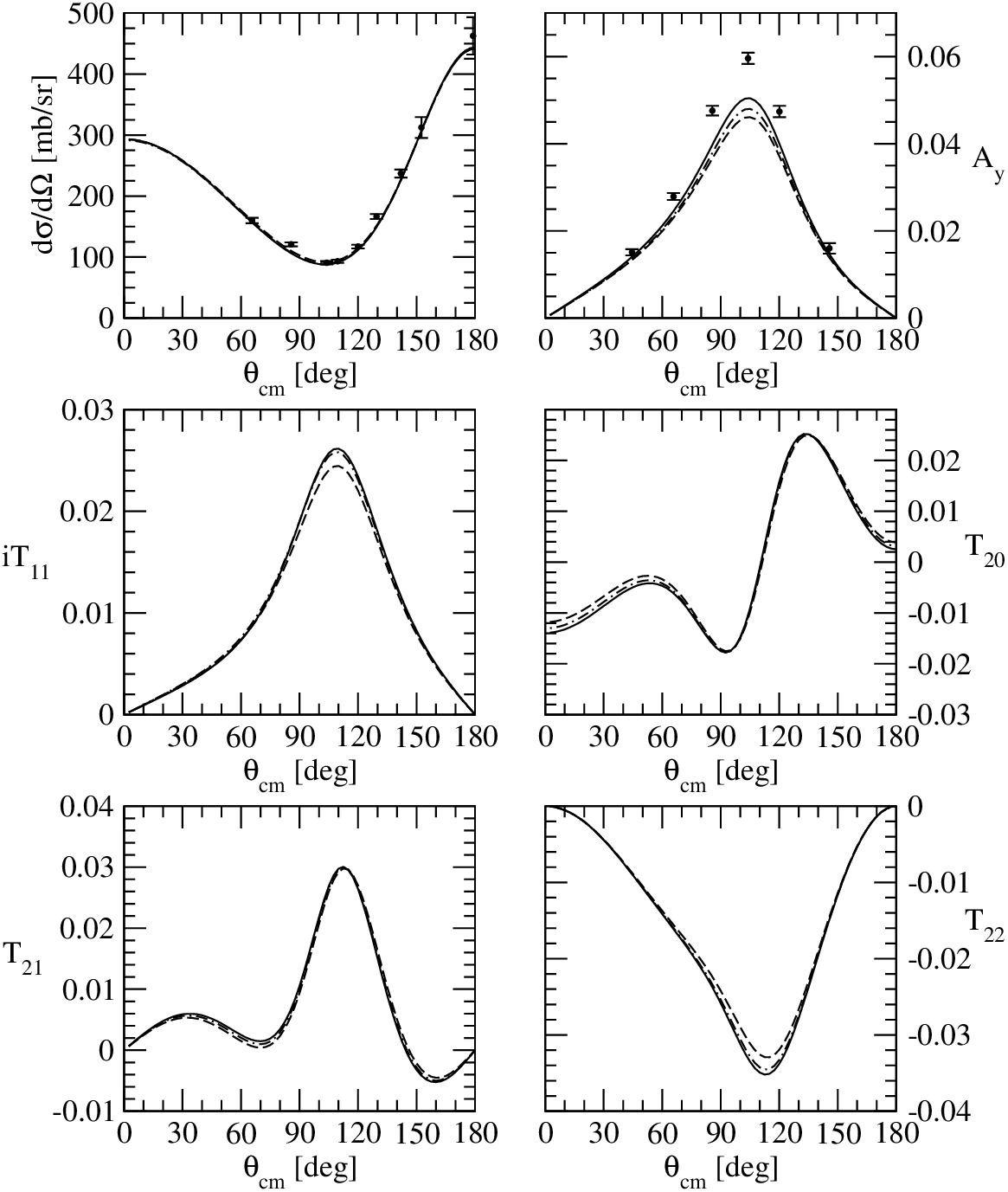}
\end{center}
\caption{\label{fig:ndobs2} 
Same as Fig.~\protect\ref{fig:ndobs1.33}, but for 
$E_{cm}=2.0$ MeV. The 
experimental data are from Ref.~\protect\cite{Sch83} for $d\sigma/d\Omega$,
and Ref.~\cite{McA94} for $A_y$. The incident neutron (deuteron) 
is $E_n=3.0$ MeV ($E_d=6.0$ MeV).}
\end{figure}

\section{Conclusions}
\label{sec:concl}

Following our previous studies on the HH method revisited to 
work in momentum-space~\cite{Viv06,Kie08}, we 
have implemented our technique to study the $N-d$ elastic scattering 
problem at center-of-mass energies below deuteron breakup threshold, 
using both local and non-local realistic nuclear interactions. 
Using this method, it is possible to accurately calculate 
$N-d$ scattering observables at very low energies, 
including the contribution from the Coulomb potential 
as well as from higher order 
electromagnetic terms, such as the MM interaction.
In particular, it is the first time that nuclear model including 
non-local two-nucleon interactions plus TNIs are used to describe $p-d$ 
scattering at very low energies.
We have studied several observables, as scattering lengths, 
differential cross section, vector and tensor analyzing powers, and 
we have compared our results with the available experimental data.
Our main conclusions can be summarized as follows: (i) the results 
obtained from the local AV18 and the non-local N3LO-Idaho two-nucleon 
interaction are quite different from each other, especially for the 
vector polarization observables $A_y$ and $iT_{11}$ in the 
maximum region. (ii) The differences between AV18 and N3LO-Idaho 
results are strongly reduced when the MM effects are included in the 
AV18 calculation. To be noticed that the MM effects are indirectly 
included 
in the nuclear N3LO-Idaho interaction, since in the fitting procedure 
for this model only the point Coulomb interaction between $pp$ is used.
(iii) Among the TNIs here considered, the N2LO model performs 
slightly better than 
the UIX one. The N3LO-Idaho/N2LO results are in fact generally
closer to the 
experimental data than the N3LO-Idaho/UIXp ones. (iv) The 
$V_{low\!-\!k}$ two-nucleon interaction model has also been considered, 
obtained from AV18 with a cutoff parameter $\Lambda=2.2$ fm$^{-1}$, 
fitted to reproduce the triton binding energy. The 
$V_{low\!-\!k}$ results for those observables 
sensitive to $S$-wave scattering, such as the scattering lengths, are in 
strong disagreement with the experimental data and quite different 
from the corresponding AV18/UIX ones. On the contrary, the results for those 
observables sensitive to $P$- and $D$-wave scattering, 
such as vector and tensor 
analyzing powers, are very similar to the corresponding 
AV18/UIX ones. 
Further studies on these aspects are currently underway.
We expect to extend the present approach to the $A=4$ 
scattering problem below breakup threshold, as already done for 
zero-energy scattering in Ref.~\cite{Kie08}.

\end{document}